\documentclass[
 reprint,
 superscriptaddress,
 amsmath,amssymb,
 prd,
floatfix,
]{revtex4-2}

\usepackage{graphicx}
\usepackage[caption=false]{subfig}
\usepackage{footnote}
\usepackage{dcolumn}
\usepackage{bm}
\usepackage{array}
\usepackage[hidelinks]{hyperref}

\usepackage[
separate-uncertainty = true,
multi-part-units=single
]{siunitx}
\usepackage{xcolor}
\usepackage[version=4]{mhchem}
\usepackage[normalem]{ulem}
\usepackage{tablefootnote}
\usepackage[capitalise]{cleveref}

\usepackage{xfrac}

\DeclareSIUnit\keVnr{keV_{nr}}
\DeclareSIUnit\keVee{keV_{ee}}

\newcommand\XeOneTwoSeven{\ce{^{127}Xe}}

\newcommand{\dquotes}[1]{``#1''}

\def\geantFour  {\mbox{\scshape Geant4}\:\:}

\begin{document}

\title{Background Determination for the LUX-ZEPLIN (LZ) Dark Matter Experiment}
\author{J.~Aalbers}
\affiliation{SLAC National Accelerator Laboratory, Menlo Park, CA 94025-7015, USA}
\affiliation{Kavli Institute for Particle Astrophysics and Cosmology, Stanford University, Stanford, CA  94305-4085 USA}

\author{D.S.~Akerib}
\affiliation{SLAC National Accelerator Laboratory, Menlo Park, CA 94025-7015, USA}
\affiliation{Kavli Institute for Particle Astrophysics and Cosmology, Stanford University, Stanford, CA  94305-4085 USA}

\author{A.K.~Al Musalhi}
\affiliation{University of Oxford, Department of Physics, Oxford OX1 3RH, UK}

\author{F.~Alder}
\affiliation{University College London (UCL), Department of Physics and Astronomy, London WC1E 6BT, UK}

\author{S.K.~Alsum}
\affiliation{University of Wisconsin-Madison, Department of Physics, Madison, WI 53706-1390, USA}

\author{C.S.~Amarasinghe}
\affiliation{University of Michigan, Randall Laboratory of Physics, Ann Arbor, MI 48109-1040, USA}

\author{A.~Ames}
\affiliation{SLAC National Accelerator Laboratory, Menlo Park, CA 94025-7015, USA}
\affiliation{Kavli Institute for Particle Astrophysics and Cosmology, Stanford University, Stanford, CA  94305-4085 USA}

\author{T.J.~Anderson}
\affiliation{SLAC National Accelerator Laboratory, Menlo Park, CA 94025-7015, USA}
\affiliation{Kavli Institute for Particle Astrophysics and Cosmology, Stanford University, Stanford, CA  94305-4085 USA}

\author{N.~Angelides}
\affiliation{University College London (UCL), Department of Physics and Astronomy, London WC1E 6BT, UK}
\affiliation{Imperial College London, Physics Department, Blackett Laboratory, London SW7 2AZ, UK}

\author{H.M.~Ara\'{u}jo}
\affiliation{Imperial College London, Physics Department, Blackett Laboratory, London SW7 2AZ, UK}

% ...

\collaboration{The LUX-ZEPLIN (LZ) Collaboration}

\date{\today}

\begin{abstract}
The LUX-ZEPLIN experiment recently reported limits on WIMP-nucleus interactions from its initial science run, down to $9.2\times10^{-48}$\,cm$^2$ for the spin-independent interaction of a \SI{36}{GeV/c\squared} WIMP at 90\% confidence level. In this paper, we present a comprehensive analysis of the backgrounds important for this result and for other upcoming physics analyses, including neutrinoless double-beta decay searches and effective field theory interpretations of LUX-ZEPLIN data. We confirm that the \textit{in-situ} determinations of bulk and fixed radioactive backgrounds are consistent with expectations from the \textit{ex-situ} assays. The observed background rate after WIMP search criteria were applied was $(6.3\pm0.5)\times10^{-5}$ events/keV$_{ee}$/kg/day in the low-energy region, approximately 60 times lower than the equivalent rate reported by the LUX experiment.
\end{abstract}

\maketitle



\section{Introduction}
\label{sec:Intro}
LUX-ZEPLIN (LZ) is an experiment optimized for the observation of signals from weakly interacting massive particles (WIMPs) having masses in excess of about 5\,GeV/c$^2$. In addition to WIMPs, LZ is sensitive to a range of other hypothetical processes as well as physics beyond the Standard Model including, but not limited to, neutrinoless double-beta ($0\nu\beta\beta$) decay \cite{LZ-Xe136-0vBB, LZ-Xe134-0vBB}, axions and axion-like particles \cite{LZ-LEER}. To maximize discovery potential for WIMPs or any other search candidates, or to set reliable upper limits on their interactions, sources that could produce similar signatures must be well-understood. 

The LZ experiment is described in detail in Refs.~\cite{LZ-CDR,LZ-TDR,LZ-detector}. The central detector is a liquid-gas xenon time projection chamber (TPC) of 7-tonne active mass, wherein particle interactions or ``events" are observed via the collection of light. Two separate signals arise from the detection of prompt scintillation and delayed electroluminescence light from a given interaction, the latter of which is created by charge extracted into the gas phase. These are known as S1 and S2, respectively. The combination of S1 and S2 can be used to reconstruct the energy and position of an event, as well as indicate the type of interaction, whether it was an electronic recoil (ER) or nuclear recoil (NR) on xenon, the target medium. The excellent, $\mathcal{O}$(mm) position resolution facilitates the identification of single scatter (SS) and multiple scatter (MS) event classifications, and offers accurate detector fiducialization \cite{LZ-SR1result}. All of these TPC features help to discriminate between potential signal and background events. In the case of WIMPs, their few to tens of keV single NRs can be well distinguished from the majority background ERs of similar energy, which typically occur towards the edges of the detector and often happen as part of MS events.

Two additional detectors form an anti-coincidence, active veto system: a 2-tonne liquid xenon “Skin” directly surrounding the TPC and a 17-tonne gadolinium-loaded liquid scintillator (GdLS) outer detector (OD). The Skin provides an extra, instrumented buffer layer between the TPC and outside radiation, leveraging the high probability that gamma rays which traverse this detector will scatter within it. The design of the OD is optimized for the tagging of neutrons, with gadolinium having an extremely high thermal neutron capture cross-section. The Skin and OD are therefore effective at tagging gamma rays and neutrons that enter or exit the TPC, which would signify events from conventional sources.

LZ operates in the Davis Cavern of the Sanford Underground Research Facility (SURF) in Lead, South Dakota, where it is well-shielded from cosmic rays by 4300 meters water equivalent (m.w.e.) rock overburden \cite{Heise_2021}. Furthermore, the entire detector configuration resides within a tank filled with 238 tonnes of ultrapure water, providing $>$\,1.2 m.w.e of passive shielding in every direction. Background events in the TPC thus result predominantly from radioactivity internal to the LZ assembly. 

An extensive radioassay campaign was undertaken to inform the material selection for the construction of the experiment, to ensure and confirm low WIMP-search background burden arising from the detector components \cite{LZ-Cleanliness}. Construction of the TPC was undertaken in a class 1000, radon-reduced cleanroom under strict cleanliness protocols to limit the plate-out of radon progeny to $<$\,0.5\,mBq/m$^2$ and the deposition of airborne dust to $<$\,500\,ng/cm$^2$ on detector surfaces. The xenon is purified \textit{in situ} via a hot zirconium getter and an inline radon reduction system \cite{Michigan-RRS}. Before underground deployment, the xenon also underwent charcoal chromatography at SLAC to reduce the Kr and Ar contamination in order to control the beta backgrounds from $^{85}$Kr and $^{39}$Ar. The expectations under these background mitigation strategies and associated requirements, as they pertain to a WIMP search, were formerly set out in our sensitivity projections in Ref.~\cite{LZ-sensitivity}.

This paper reports the current understanding of our backgrounds from inferences made with LZ’s initial science run (SR1); 89 live days of data taken between December 2021 and May 2022. It details the general background observations and how their determinations compare to previous \textit{ex-situ} estimates, before motivating how they affected the physics conclusions of the experiment’s first WIMP search results in Ref.~\cite{LZ-SR1result}. Section~\ref{sec:Sims} discusses the simulations and modeling that underpin the latest calculations of our expected background events. Section~\ref{sec:TPCBackgroundFits} covers measurements of our backgrounds from examining distributions in energy space (external gamma-ray and beta radiations; electron captures from noble radioisotopes within the xenon; alpha particles from radon and its daughters). Section~\ref{sec:CountedTPCBackgrounds} documents backgrounds which can be counted within our data (muons and neutrons). Section~\ref{sec:WIMPBackgrounds} explains how these backgrounds impacted our first WIMP analysis, as well as describes our physics backgrounds informed by other experiments and calculations ($^{136}$Xe two-neutrino (2$\nu\beta\beta$) double-beta decay, $^{124}$Xe two-neutrino double-electron capture (2$\nu$DEC), solar and atmospheric neutrinos), sources specific to such low-energy searches ($^{37}$Ar, accidental coincidences and wall backgrounds) and the choice of the backgrounds-driven fiducial volume (FV) for SR1. Section~\ref{sec:Conclusions} provides a conclusion.

\section{Background Simulations}
\label{sec:Sims}
Detailed Monte Carlo simulations employing two in-house software packages were used to estimate the background contributions from detector materials, xenon contaminants, and the laboratory environment. The first package, BACCARAT \cite{LZSims}, contains the LZ detector geometry, which has seen updates since the sensitivity studies to closely match the as-built detectors. It tracks particles using \geantFour \cite{GEANT4} and identifies their interaction points in the detectors. Energy depositions were recorded and passed to the second package, LZLAMA (LZ Light Analysis Montecarlo Application). LZLAMA models the detector response based on NEST (Noble Element Simulation Technique) \cite{NEST1, NEST2} and returns observables such as S1 and S2 pulse size, in photons detected (phd), timing, and reconstructed event location.

The TPC detector response model in LZLAMA was tuned with tritium calibration data and verified with deuterium-deuterium (DD) neutron generator data, with particular focus on response matching for low-energy events pertinent to the WIMP search. Among the important parameters tuned for SR1 were the photon gain, $g1= 0.1136\pm0.0020$ phd/photon, and the charge gain, $g2= 47.07\pm1.13$ phd/electron. The full set of SR1 detector parameters that were provided to NEST can be found in Ref.~\cite{LZ-supplementalMaterial}. A uniform electric field strength of 193\,V/cm was used for calculating light and charge yields, as well as the electron-ion recombination probability. 

Non-uniformity in the field near the TPC walls leads to curvature of the drift paths of the ionization electrons. This was modeled in LZLAMA using a drift map as described in Ref.~\cite{LZSims}, tuned so that the mean reconstructed position of the wall as a function of drift time matched that of an injected $^{83m}$Kr calibration source. Finite resolution in position reconstruction was incorporated via a model that was similarly constrained using the position variation from events at the wall. 

Pulses that appear close together in time may be reconstructed as a single pulse. This pulse merging was approximated in simulation using two independent models for the merging of S1 and S2 pulses, both of which were dependent on relative pulse time, pulse size, and interaction position. $^{83m}$Kr calibration events were used to tune the S1 model, given the high likelihood of producing S1 pulses close together in time, due to the two-step structure of its decay. AmLi calibration events were chosen for the S2 model tuning, given that they often consist of multiple interactions in close proximity where S2s could merge. The tunings of both models achieved a better than 5\% agreement when compared to these respective calibrations. Simulated background events were found to be far less likely than $^{83m}$Kr or AmLi events to contain multiple interactions close enough in space and/or time to be edge cases in the pulse merging implementation. This suggested that the fraction of events producing SSs, or the SS efficiency, should be faithfully reproduced in simulations. Indeed, the consistency of fits to background data conducted with simulation outputs, as discussed in Section~\ref{sec:bgfitting}, implied good agreement in the SS efficiency across all energies.

The background simulations include 1144 detector volume and radioisotope pairings. Results from a comprehensive radioassay and screening campaign of all detector materials in  Ref.~\cite{LZ-Cleanliness} informed these fixed contaminants. The $^{238}$U chain was broken into “early” and “late” parts, separated by $^{226}$Ra, since its long half-life of 1600 years would delay the re-establishment of secular equilibrium within the chain following chemical processing of the materials. $^{232}$Th was similarly split into early and late chains at the level of $^{224}$Ra.

The simulations were designed to be as extensive and as granular as possible. However, more materials were assayed than were represented in the BACCARAT geometry; the smaller components by mass and size, for which there was no direct corresponding volume in the simulations, were proxied by nearby larger components. The simulation statistics of these larger components and their initial normalizations, as assumed for background fitting, were accordingly scaled to account for the estimated activities of these smaller volumes. 

Gamma rays from the cavern rock can penetrate the water tank and irradiate the TPC to constitute an external background. The simulation here relied on the multi-staged process detailed in Ref.~\cite{LZSims}, and the initial normalizations on the \textit{in-situ} measurements described in Ref.~\cite{LZ-CavernGammas}. Additionally, a range of xenon contaminants, activation lines, and neutrino fluxes were considered.

\section{Global TPC Background Determination}
\label{sec:TPCBackgroundFits}
\label{sec:bgfitting}
To provide a global overview of the backgrounds present in SR1, the full dataset was analyzed in reconstructed energy space to identify and constrain features particular to certain backgrounds. These include mono-energetic peaks from total gamma-ray or alpha energy depositions, as well as more complicated spectra with superposed beta and gamma-ray contributions. Fits were either conducted with template shapes to these spectral features, or with spectra simulated using the framework outlined in Section~\ref{sec:Sims}.

The majority of these fits were conducted with SS events, as this was the classification focused on for the SR1 WIMP search and consequently the one for which the event reconstruction and analysis cut efficiencies had been well vetted \cite{LZ-SR1result}. SS event energies were reconstructed using

\begin{equation}
\label{eq:energyRecon}
E = W \left( \frac{\text{S1}c}{g1} + \frac{\text{S2}c_{\text{bot}}}{g2_{\text{bot}}} \right),
\end{equation}

where a $W$ value of 13.5\,eV \cite{NEST-MDPI} was assumed, the $c$ denotes the S1 and S2 pulse sizes have been corrected for position, via the processes described in Ref.~\cite{LZ-SR1result}, and the subscript ``bot" signifies the quantity was derived using only light collected in the bottom photomultiplier tube (PMT) array. $g1$ was as reported in Section~\ref{sec:Sims} and $g2_{\text{bot}} = 14.89\pm0.48$ phd/electron, the bottom array-only equivalent of the charge gain. Bottom array-only S2s were used since, when reconstructing events with energy of $\mathcal{O}$(100\,keV) and higher, the large amount of light seen in the top array can cause amplifier saturation, which in turn leads to an under-reporting of the S2 pulse size. 

For fits that were particularly position-sensitive, extra corrections were employed for the S1s and S2s to provide an adequately homogenized response across the whole detector for improved energy resolution. Correction factors were derived by fitting the position dependence of the S1 and S2 signals produced by 5.49\,MeV alpha particles from $^{222}$Rn decays in the active TPC volume. The resultant correction functions, smooth in all three position coordinates, are henceforth referred to as $^{222}$Rn alpha-based corrections. These correction factors were utilized for the subset of radon-chain alpha studies that used SS events, and for the model fitting of higher-energy detector gamma-ray radiation, where there was a need to ensure sufficient energy resolution up to the 2.6\,MeV $^{208}$Tl line.

The xenon activation peak analysis in Section~\ref{sec:XeActivation} was pursued with both SS and MS events. Since energy resolution was not a concern for this particular study, the inclusion of MS events was justified as increased statistics were required for accurate event counting during later time periods of SR1, when xenon activation isotopes had largely decayed. In the case of MS events, in which an S1 is paired with two or more S2s, Equation~\ref{eq:energyRecon} instead used the summed total of the S2$c$ pulse sizes, and the S1 was corrected by calculating the position correction weighted by the pulse sizes of each S2.

Exceptions to Equation~\ref{eq:energyRecon} were made for a selection of the radon-chain alpha analyses and the cavern gamma-ray studies. In these cases, energy was calculated using S1 information only. This was necessary for the latter analysis since it was performed in a xenon gas environment prior to the application of electric fields, when no S2 signals were available. For the radon-chain alpha studies, S2 information was often difficult to reconstruct due to the interaction location or topology, and many events of interest were not classified as SS or MS. In this case, an S1-derived energy scale was found to provide adequate resolution for alpha events, given the large S1s associated with the $\mathcal{O}$(MeV) alpha energy depositions. 

Different energy ranges and volumes were explored to best constrain specific backgrounds. The radon alpha peaks were examined in an otherwise background-poor region above 3\,MeV using both S1-only and S1+S2-based analyses (Section~\ref{sec:RadonAlphas}). Separately, the identifiable xenon activation peaks were isolated in an intermediate energy range of 200--450\,keV (Section~\ref{sec:XeActivation}). These and internal backgrounds, including the beta spectra from $^{214}$Pb in the $^{222}$Rn chain and the $^{212}$Pb from the $^{220}$Rn chain, whose rates were informed by the radon alpha fits, were then constrained in an inner one-tonne volume in the 80--700\,keV range (Section~\ref{sec:BetaFit}). Gamma-ray backgrounds were examined in commissioning data, when cavern gamma rays were expected to be dominant (Section~\ref{sec:cavernGammaRayMeasurement}). The gamma-ray contributions were then fitted with SR1 data in the 1--2.7\,MeV region (Section~\ref{sec:SR1GammaRays}), before a final global fit in the SR1 FV was attempted, incorporating the internal backgrounds, spanning 80--2700\,keV (Section~\ref{sec:SR1FVFit}).

\subsection{The Radon Alpha Region ($>$3\,MeV)}
\label{sec:RadonAlphas}

\begin{figure}[!t]
    \centering
    \includegraphics[width=\columnwidth]{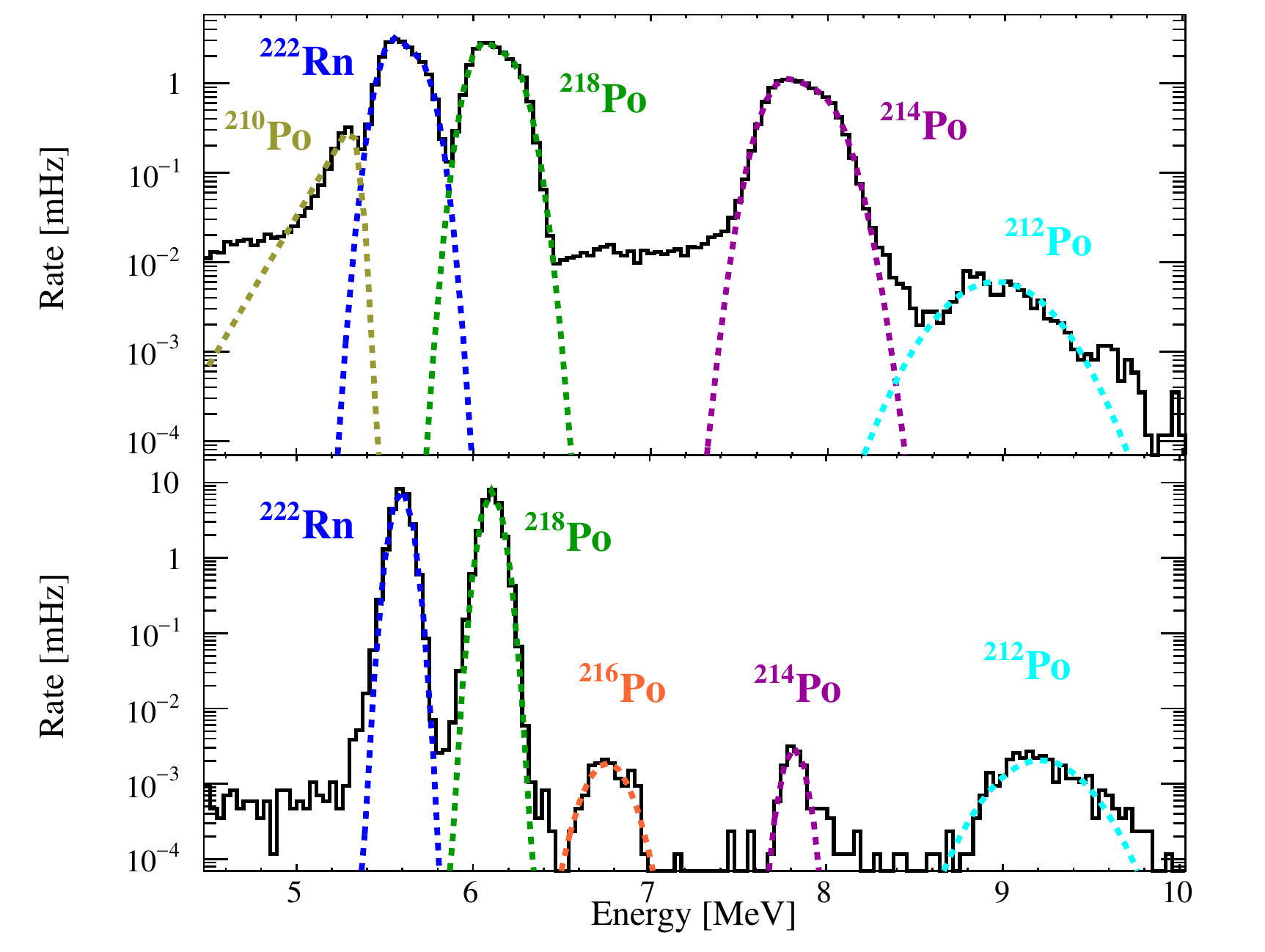}
    \centering
    \caption{Fitted radon alpha spectra in S1-linear calibrated energy. \emph{Top:} All alpha events in the TPC are shown with only TBA cuts applied to remove excess grid and below-cathode alpha events. The $^{222}$Rn, $^{218}$Po, and $^{214}$Po were modeled as double Gaussians, $^{212}$Po as a single Gaussian, and $^{210}$Po as a modified Crystal Ball function. \emph{Bottom:} Alpha spectra for SS events in the SR1 fiducial volume (see Section~\ref{sec:WallFV}), where position information from the S2 pulse was used to correct the spectra, resulting in improved resolution. The sum of five Gaussians was used to model the $^{222}$Rn, $^{218}$Po, $^{216}$Po, $^{214}$Po, and $^{212}$Po peaks.}
    \label{fig:RadonAlphaPeaks}
\end{figure}

$^{222}$Rn and $^{220}$Rn emanate from primordial $^{238}$U and $^{232}$Th decay chains present in detector materials and dust, dispersing within the active liquid xenon volume. Radon and many of its progeny provide the only sources of alpha particles in the TPC. These alpha particles are highly energetic, producing $\mathcal{O}$(MeV) energy depositions; they are also densely ionizing in liquid xenon, leading to high recombination yields for their interactions. These two facts mean that radon alphas produce extremely large and readily identifiable S1 signals, far removed from those of other backgrounds.

The analyses that follow rely on these S1s for investigations of the Rn progeny. Two separate fits were performed on S1 pulse sizes, each with different motivations and merits: one which was agnostic to the classification of the alpha event, and another which used solely events classified as SS. 

The classification-agnostic fit had the advantage of being sensitive to more populations than the SS-only fit since, for many alpha decays, the event may not be SS-classified. For example, alpha decays that occur on or near the wall may lose much of their S2 signal due to electron attachment on the polytetrafluoroethylene (PTFE), leading to misclassification. Bi-Po events, in which both the $^{214}$Bi ($^{212}$Bi) and $^{214}$Po ($^{212}$Po) decays occur within the event window, are complicated due to the possibilities of overlapping signals from the beta and alpha decays, and multiple scatters from the gamma-ray daughters of the Bi decay. In these events, the top-bottom asymmetry (TBA) of the S1 signal, i.e. the ratio of the difference between S1 light collected by the top and bottom PMT arrays to the total S1 light collected, can be leveraged. Regardless of event classification, alphas produce distinct bands in S1 area-TBA space, where the size of the signal is primarily determined by the light collection efficiency as a function of depth in the detector. 

\begin{figure*}[!t]
\centering
 \subfloat[][Observed $^{218}$Po Distribution]
 {\includegraphics[width=0.45\textwidth]{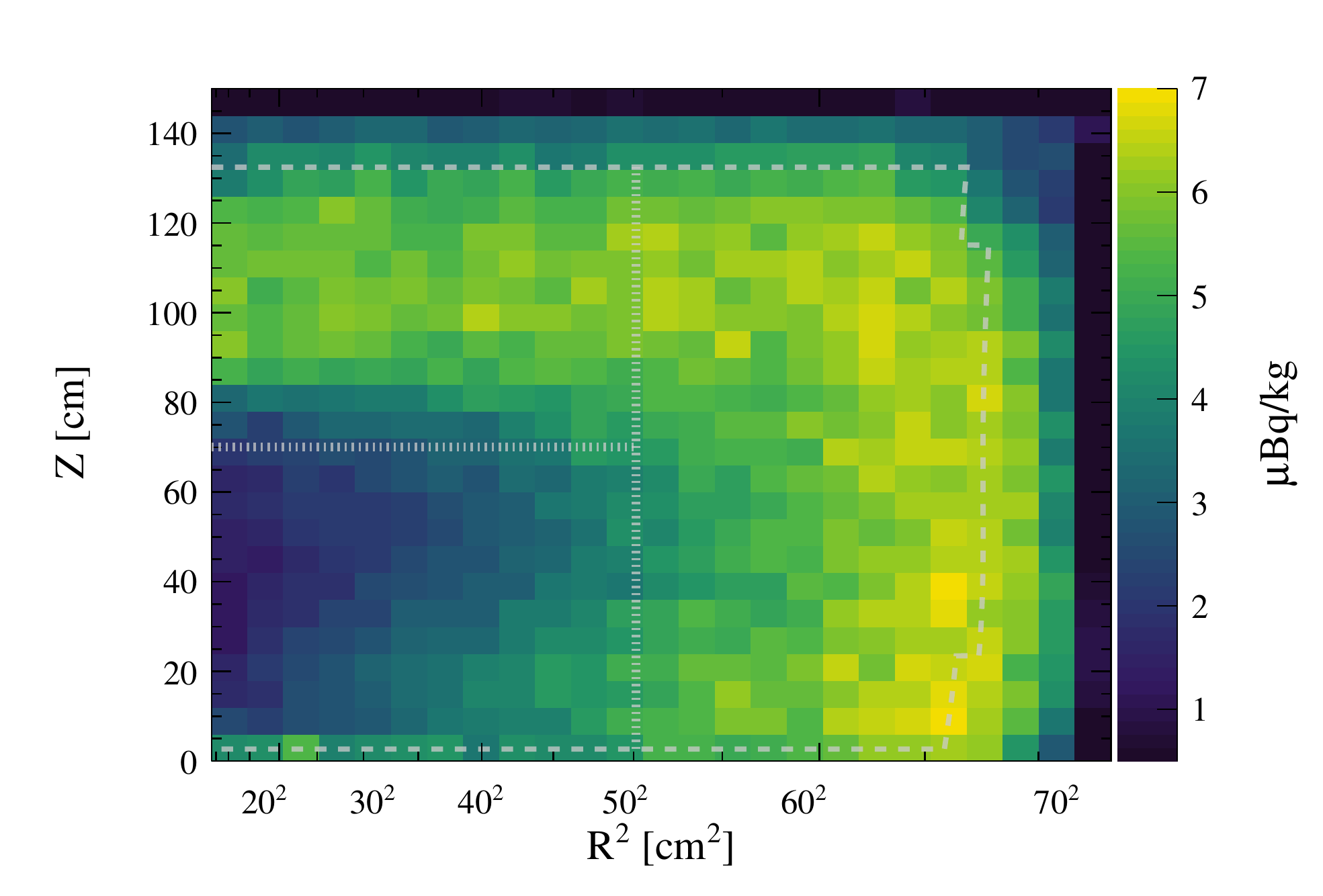} \label{fig:Radon_positions_fig_a}}
 \qquad
 \subfloat[][Observed $^{214}$Po Distribution]
 {\includegraphics[width=0.45\textwidth]{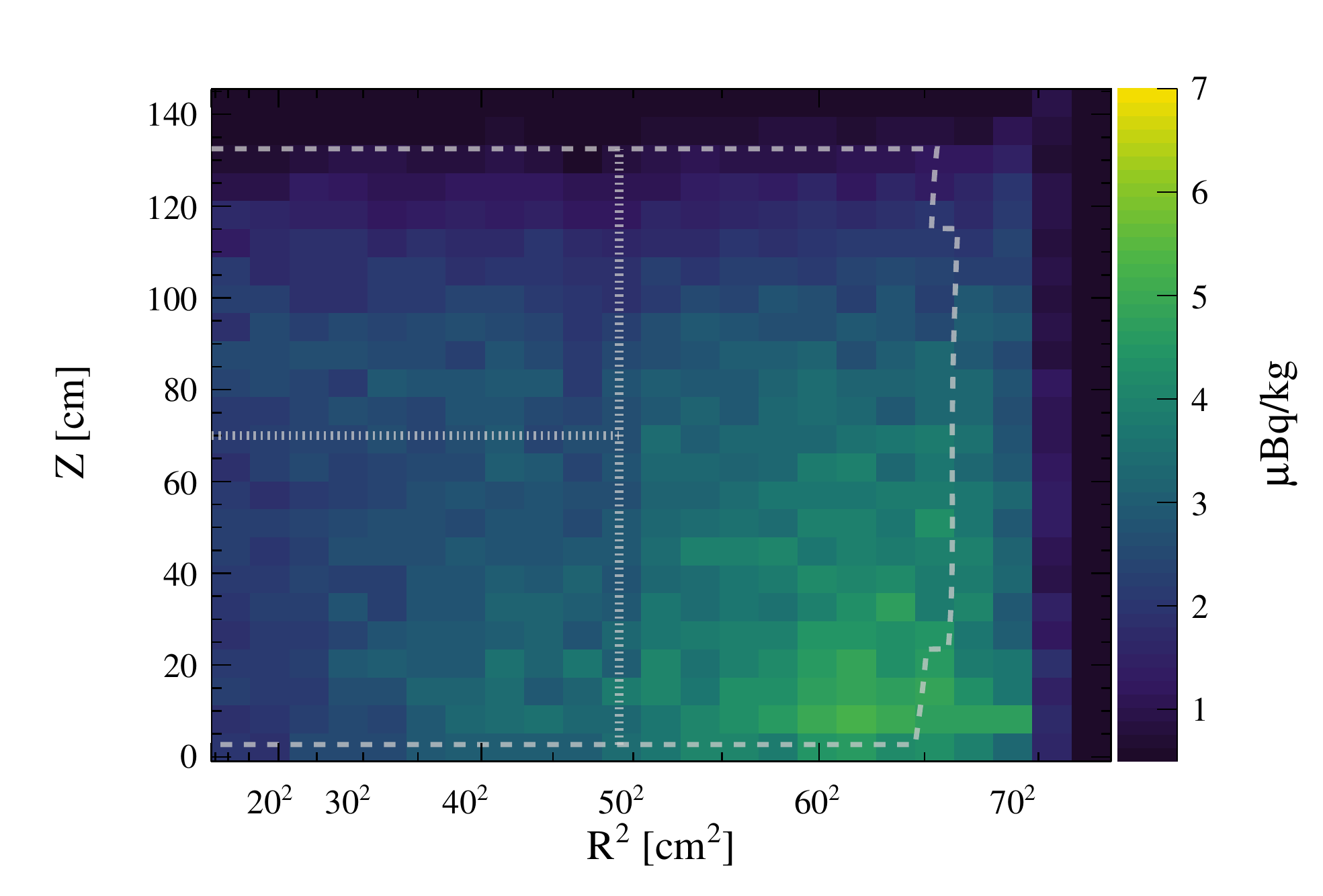} \label{fig:Radon_positions_fig_b}}
 \qquad
 \subfloat[][Simulated $^{214}$Pb Distribution]
 {\includegraphics[width=0.45\textwidth]{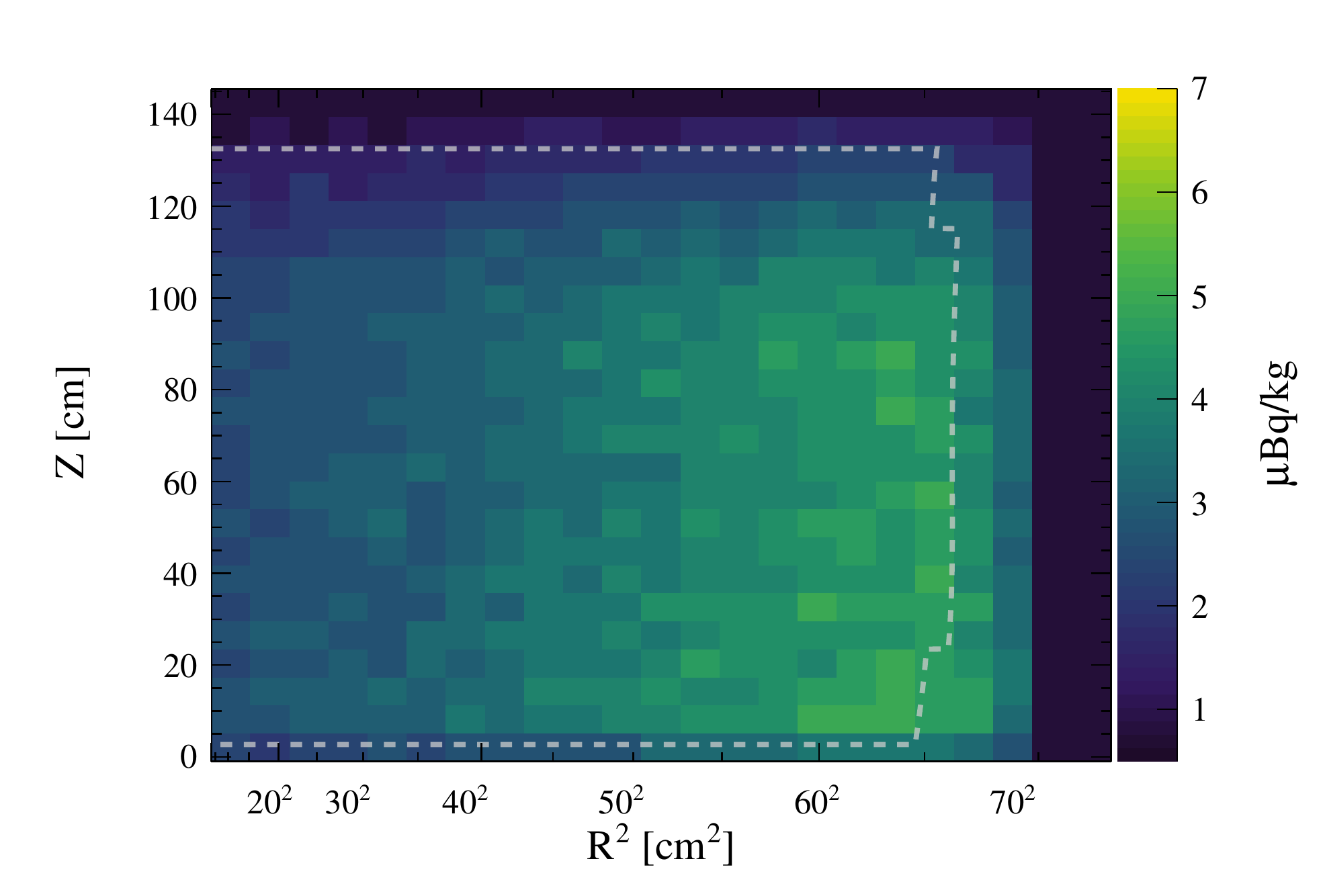} \label{fig:Radon_positions_fig_c}}
 \qquad
 \subfloat[][Simulated $^{214}$Po Distribution]
 {\includegraphics[width=0.45\textwidth]{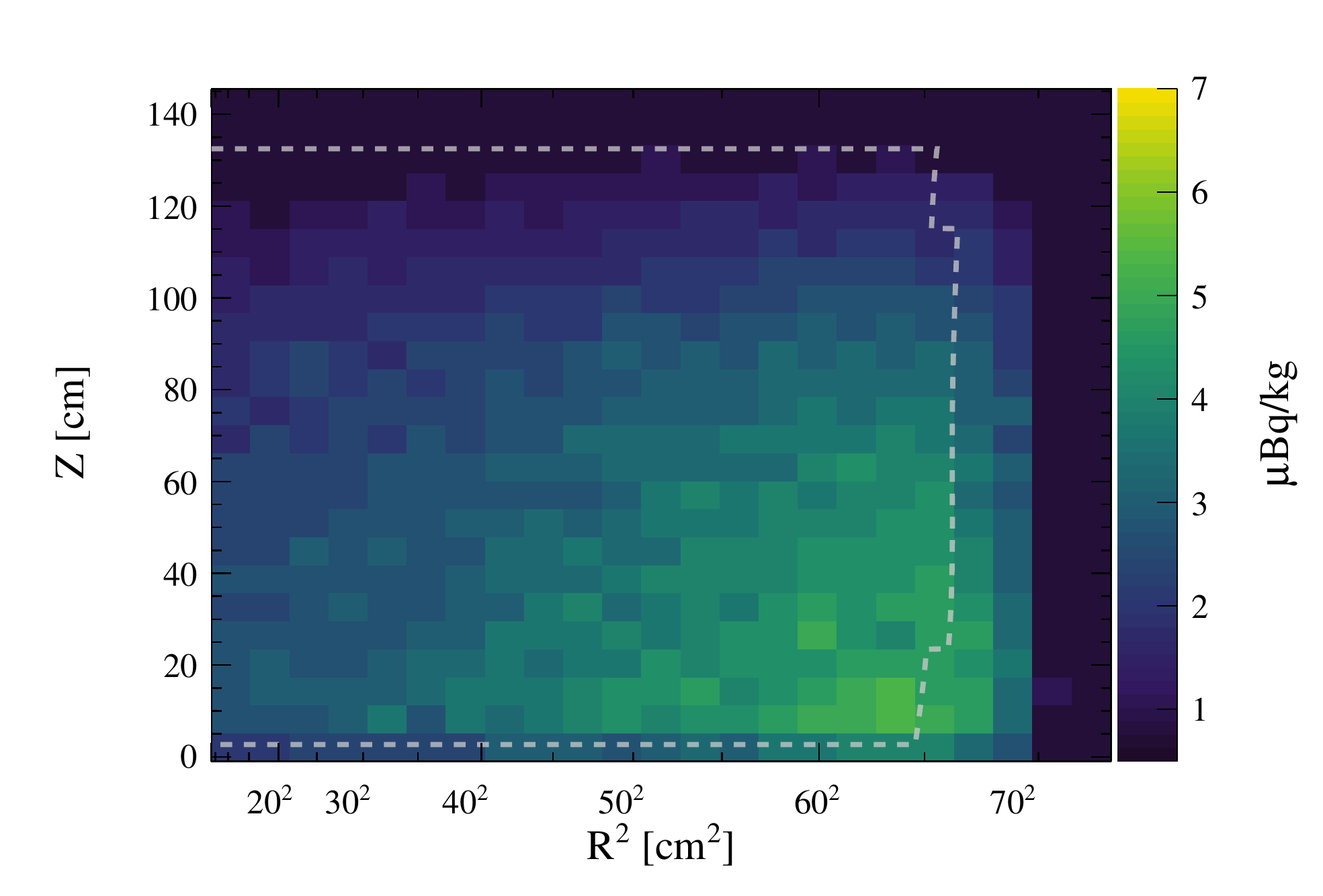} \label{fig:Radon_positions_fig_d}}
\caption{The observed and simulated distributions of selected $^{222}$Rn daughters. The $^{218}$Po distribution in data was found from SS events and is shown in panel (a); the $^{222}$Rn distribution can be considered to be nearly identical to that of $^{218}$Po. The measured $^{214}$Po distribution is shown in panel (b), for which the S1-S2 pair of $^{214}$Po was extracted from each Bi-Po event to reconstruct its position analogously to a SS event. Simulated $^{214}$Pb and $^{214}$Po distributions in panels (c) and (d), respectively. The robustness of the toy model was validated from the agreement between the simulated and observed $^{214}$Po distributions. The FV boundary is shown with a dashed gray line on all panels, whereas for the observed distributions, two additional dotted lines further separate the FV into upper, lower, and outer regions. These volumes are used in Table~\ref{tab:Radon_AlphaActivities} to quantify the non-uniformity of the observed alphas.}
\label{fig:Radon_positions}
\end{figure*}

In order to measure alpha rates in the full TPC, the S1 pulse sizes were first corrected for the depth-dependent light collection efficiency. A common third-degree polynomial was fitted to the alpha S1 area-TBA bands, which was used to normalize the S1 responses to the vertical center of the detector. A TBA cut was applied to remove events from the grids and below the cathode, and all S1s above 30,000\,phd were analyzed. The resultant alpha distributions show good distinction between the different alpha energies, as seen in Figure~\ref{fig:RadonAlphaPeaks}, allowing effective identification of each population with very small contamination from neighboring peaks. These alpha peaks were simultaneously fit, with $^{222}$Rn, $^{218}$Po, and $^{214}$Po modeled as the sum of double Gaussians, and $^{212}$Po as a single Gaussian. Double Gaussians were used to account for the observed skewness of some of the peaks as a result of the radially-dependent light collection efficiency. The double Gaussian model for the observed distributions was found to capture well the combination of high radon and radon daughter concentrations near the TPC wall, with the abrupt changes in radial profiles as seen in Figure~\ref{fig:Radon_positions}, and the impact of proximity to the wall on the light collection efficiency. The $^{212}$Po alphas were often merged with $^{212}$Bi betas, thus slightly skewing the peak to higher energies. However, given the low rate of $^{212}$Po, it was found a Gaussian fit worked sufficiently well to capture these $^{212}$Po events. $^{210}$Po was fit with a modified Crystal Ball function --- a Gaussian core with a low-energy exponential tail as defined in Ref.~\cite{Das:2016stf} --- to account for alphas that lose energy before interacting in the xenon, having come from decays embedded in the TPC PTFE walls, and to account for events on the wall surfaces that suffer from poor S1 light collection. 

The classification-agnostic fit is illustrated in the top panel of Figure~\ref{fig:RadonAlphaPeaks}. The results were used to populate the “TPC Rate” column of Table~\ref{tab:Radon_AlphaActivities}, having been normalized by the mass of the fit volume, assuming it to be representative of the whole TPC. $^{216}$Po and $^{220}$Rn were not included in this analysis as $^{216}$Po was not well-resolved in this space, and the $^{220}$Rn contribution, which overlaps $^{218}$Po, was too subdominant to be identified. 
     
In the second, SS-only fit, S2 information could be exploited to more accurately inform the position of the interaction. Whilst significant alpha populations that were not classified as SS were lost from the analysis, a better correction of the pulse size as a function of position could be implemented for those still present. The $^{222}$Rn alpha-based corrections discussed in the introduction to Section~\ref{sec:TPCBackgroundFits} was employed for the alpha bands to correct the S1 size to the vertical and radial center of the TPC. This improved the resolution of the alpha peaks and revealed the presence of $^{216}$Po, expected to manifest as a SS the majority of time, that was previously obstructed by alphas from other classifications. The alpha decay SS spectra in the SR1 FV (illustrated in Figure~\ref{fig:SS_DetER_FitVolumes} and defined in Section~\ref{sec:WallFV}) are shown in the bottom panel of Figure~\ref{fig:RadonAlphaPeaks}, where a simultaneous fit with five single Gaussians was performed.

To be able to use the results of the SS-only fit to quote activities for each radon-chain radioisotope, the chance of its alpha events being classified as SS had to be taken into account. The SS efficiency was determined by identifying and fitting the given alpha band in S1-TBA space separately for each classification type, and then calculating the fraction of total events that were SS. The obtained values are listed in the ``Single Scatter Efficiency" column of Table~\ref{tab:Radon_AlphaActivities}. Less than half of $^{212}$Po, and a small fraction of $^{214}$Po, was observed as SS due to the nearly instantaneous Bi-Po decay that produces overlapping signals. 

Investigation of the SS events revealed that the position distributions of $^{222}$Rn and its progeny in the TPC were highly non-uniform. The degree of inhomogeneity for $^{218}$Po and $^{214}$Po in the data can be seen in Figure~\ref{fig:Radon_positions}, panels (a) and (b), respectively. To quantify the level of anisotropy, the SS-based fits were performed in each of the lower, upper and outer sub-volumes marked out in these sub-figures, with the results detailed in the last three columns of Table~\ref{tab:Radon_AlphaActivities}.

\begin{table*}[!ht]
    \renewcommand{\arraystretch}{1.5}
    \setlength{\tabcolsep}{3pt}
    \centering
    \caption{Table of measured alpha activities. Lower, upper, and outer sub-volumes are defined as shown in Figure \ref{fig:Radon_positions}. Fits in the FV and its sub-volumes were performed with SS events and the rates are scaled to be representative of the full population using the listed SS efficiency factor. *The TPC value for $^{216}$Po was extrapolated from the SS fits performed in the FV, dividing by the volume ratio of the FV to the total drift field region of the TPC.}
     \begin{tabular}{l c c c c c c } 
        \hline \hline
         Radon  & TPC Rate & Single Scatter &  FV Rate & Lower Rate & Upper Rate & Outer Rate \\
         Isotope & [\SI{}{\micro\becquerel/\kilogram}]   & Efficiency & [\SI{}{\micro\becquerel/\kilogram}] & [\SI{}{\micro\becquerel/\kilogram}] & [\SI{}{\micro\becquerel/\kilogram}]& [\SI{}{\micro\becquerel/\kilogram}]  \\ 
         \hline
         $^{222}$Rn & 4.78 $\pm$ 0.33 & 0.96 $\pm$ 0.03  &  4.62 $\pm$ 0.87 &  2.64 $\pm$ 0.60 &  3.38 $\pm$ 0.76  &  6.32 $\pm$ 1.33  \\
         $^{218}$Po & 4.82 $\pm$ 0.34  & 0.98 $\pm$ 0.03 & 4.53 $\pm$ 0.84 &  2.64 $\pm$ 0.60 &  3.69 $\pm$ 0.83 &   6.26 $\pm$ 1.31 \\
         $^{216}$Po & *(8.2 $\pm$ 0.6)$\cdot$10$^{-3}$ & 0.56 $\pm$ 0.25 & (4.69 $\pm$ 3.15)$\cdot$10$^{-3}$ &  (6.43 $\pm$ 4.39)$\cdot$10$^{-4}$  &  (2.48 $\pm$ 1.69)$\cdot$10$^{-3}$ &  (7.63 $\pm$ 5.18)$\cdot$10$^{-3}$ \\
         $^{214}$Po & 2.65 $\pm$ 0.19 & (1.14 $\pm$ 0.38)$\cdot$10$^{-3}$  &  2.07 $\pm$ 0.95   &  0.76 $\pm$ 0.45   &   1.09 $\pm$ 0.65  &  3.37 $\pm$ 1.99 \\
         $^{212}$Po & (3.7 $\pm$ 0.3)$\cdot$10$^{-2}$  & 0.34 $\pm$ 0.08 & (1.49 $\pm$ 0.73)$\cdot$10$^{-2}$    &  (5.72 $\pm$ 2.89)$\cdot$10$^{-3}$   &    (1.43 $\pm$ 0.72)$\cdot$10$^{-2}$  & (2.89 $\pm$ 1.44)$\cdot$10$^{-2}$ \\ 
         \hline \hline
     \end{tabular}
     \label{tab:Radon_AlphaActivities}
     \end{table*}

The non-uniformity was the result of thermodynamic conditions and xenon flow in the TPC. A ``slow-mixing" inner region, which experiences a lower event rate from radon-chain isotopes than the rest of the detector, was attributed to the fact that the xenon mixing timescale was longer than the half-life of $^{222}$Rn. The observed distribution of $^{222}$Rn alpha events was very similar to the $^{218}$Po illustrated in Figure~\ref{fig:Radon_positions}a, with the two essentially identical with this bin resolution, since there is only a 3.1 minute half-life between the two decays. This can be contrasted to the distributions of later radioisotopes in the $^{222}$Rn chain, such as that of $^{214}$Po (Figure~\ref{fig:Radon_positions}b). The tendency towards the bottom of the TPC was due to charged ion movement; radon-chain decays often produce positively charged progeny that drift toward the cathode under the influence of the drift electric field \cite{EXO-200chargedions}. This charged progeny motion, coupled with the extant inhomogeneity of the parent nuclei, defines the observed position distribution. 

The incidence of charged progeny and their movements were characterized through the studies of $^{222}$Rn-$^{218}$Po decay pairs. The observed decay pairs form vectors describing the motion of $^{218}$Po, from its production to its decay. Neutral $^{218}$Po progeny move slowly with the liquid, whereas positively charged $^{218}$Po progeny move more quickly, primarily along the drift field and towards the cathode.

The charged progeny fraction of $0.46\pm0.04$ was calculated as the ratio of pairs in which the $^{218}$Po was observed to drift downward to all $^{222}$Rn-$^{218}$Po pairs, and was observed to have negligible radial dependence. The ion mobility of $0.242\pm 0.031$\,cm$^{2}$/(kV$\cdot$s) was determined from the average velocity, having fitted to the distribution of observed velocities along the field, considering the strength of the drift field and the liquid xenon density. These measured values for the charged progeny fraction and ion mobility agree to within one sigma of those reported by EXO-200 \cite{EXO-200chargedions}. 

Toy Monte Carlo mobility simulations using these values were created to understand the position distribution of sequential radon-chain decays. Starting $^{218}$Po positions were sampled from the distribution in Figure~\ref{fig:Radon_positions}a, and each subsequent ion was allowed to drift given the assumed mobility and charged fraction probability. For these simulations, convective flow was ignored as it was subdominant to charged progeny mobility, as inferred from the observed lack of movement of the neutral fraction of $^{218}$Po in our studies. The simulated $^{214}$Po distribution is shown in Figure~\ref{fig:Radon_positions}d. The drift model was validated against the observed $^{214}$Po distribution in Figure~\ref{fig:Radon_positions}b, with the activity of simulated $^{214}$Po within the fiducial volume found to be within 2$\%$ of the measured value summarized in Table ~\ref{tab:Radon_AlphaActivities}. Given the good agreement between simulations and data, the toy mobility model was used to generate an \textit{a priori} estimate for the position distribution of $^{214}$Pb (Figure~\ref{fig:Radon_positions}c). This was used to evaluate the rate of $^{214}$Pb decays in the FV, a major background for the WIMP search (Section~\ref{sec:Pb214}).

The $^{222}$Rn activities found in these studies can be compared to \textit{ex-situ} estimates compiled during the construction phase of the experiment. In sensitivity projections, the $^{222}$Rn decay rate was estimated to be \SI{1.8}{\micro\becquerel/\kilogram}, or equivalently a total of 12.6\,mBq in the active xenon volume. To determine this value, a combination of room temperature emanation measurements for several components and literature values for components that had not been screened at the time of publication were used \cite{LZ-sensitivity}. Assumptions were made for this calculation about the final detector surface cleanliness, the performance of the inline gaseous radon reduction system, and the expected reduction of radon emanation from some materials and dust particles at cryogenic versus room temperature. 

One hypothesis for the tension between the original estimate for $^{222}$Rn activity and the values reported here, given the observed position distribution of the $^{222}$Rn, is that the measured excess originates from the titanium cryostat. During the integration of the TPC, radon emanation studies of the partially assembled detector were conducted, as described in Ref.~\cite{LRT2022SDSMT}, from which it was deduced that the cryostat contributes $17.2\pm4.4$\,mBq. Possible explanations for this high emanation rate include contamination near or on the surface of the titanium, introduced during intensive assembly and integration activities; that radon has a much larger diffusion length in titanium than is typically seen in metals; or that, rather than being uniformly distributed throughout the titanium, radium has been concentrated near surfaces, which would elevate the emanation rate into the liquid xenon-filled cryostat from radium-decay radon recoils~\cite{Lindemann}. The relative contributions of these effects to the \textit{in-situ} measurement reported here, and emanation from titanium in general, requires further investigation. 

The measured $^{210}$Po rate can also be compared to that from LZ's design goals for radon daughter plate-out and surface contamination. During the fabrication and assembly of detector components, $^{222}$Rn daughters could plate-out on detector surfaces, leading to long-lived $^{210}$Pb on (and embedded in) the TPC PTFE walls. $^{210}$Po decays can pose a low-energy NR background for the WIMP search when the alpha is emitted into the PTFE, leaving a recoiling $^{206}$Pb nucleus to deposit $\mathcal{O}$(100 keV$_{\text{nr}}$) energy in the active xenon volume. These events can largely be removed via fiducialization (Section~\ref{sec:WallFV}). Nevertheless, LZ instituted a target plate-out rate for $^{210}$Po of $\leq0.5$\,mBq/m$^2$ on the TPC walls and $\leq10$\,mBq/m$^2$ on other surfaces \cite{LZ-TDR}. Using the previously discussed fit of a modified Crystal Ball function, as shown in the top panel of Figure~\ref{fig:RadonAlphaPeaks}, the measured $^{210}$Po rate was 2.32\,$\pm$\,0.15\,mBq in the TPC. This corresponds to an upper limit of $0.35\pm0.02$\,mBq/m$^2$ on the TPC walls, assuming this to be the sole origin of the $^{210}$Po, well below the design requirement.

\subsection{The Xenon Activation Region (200--450\,keV)}
\label{sec:XeActivation}

Cosmic ray-induced activation of the xenon occurs during its storage on the surface and transportation to the experimental site. The suppression of atmospheric particle fluxes by the rock overburden renders further cosmogenic activation underground insignificant. However, given the time between the last delivery of xenon to SURF (31 August 2021) and the start of science data-taking (23 December 2021) in relation to the half-lives of the activation products, these radioisotopes were expected to be prevalent in SR1. Moreover, neutron calibrations that were carried out before and during the science run resulted in further expected activation of the xenon. 

 \begin{figure}[!t]
    \centering
    \includegraphics[width=\columnwidth]{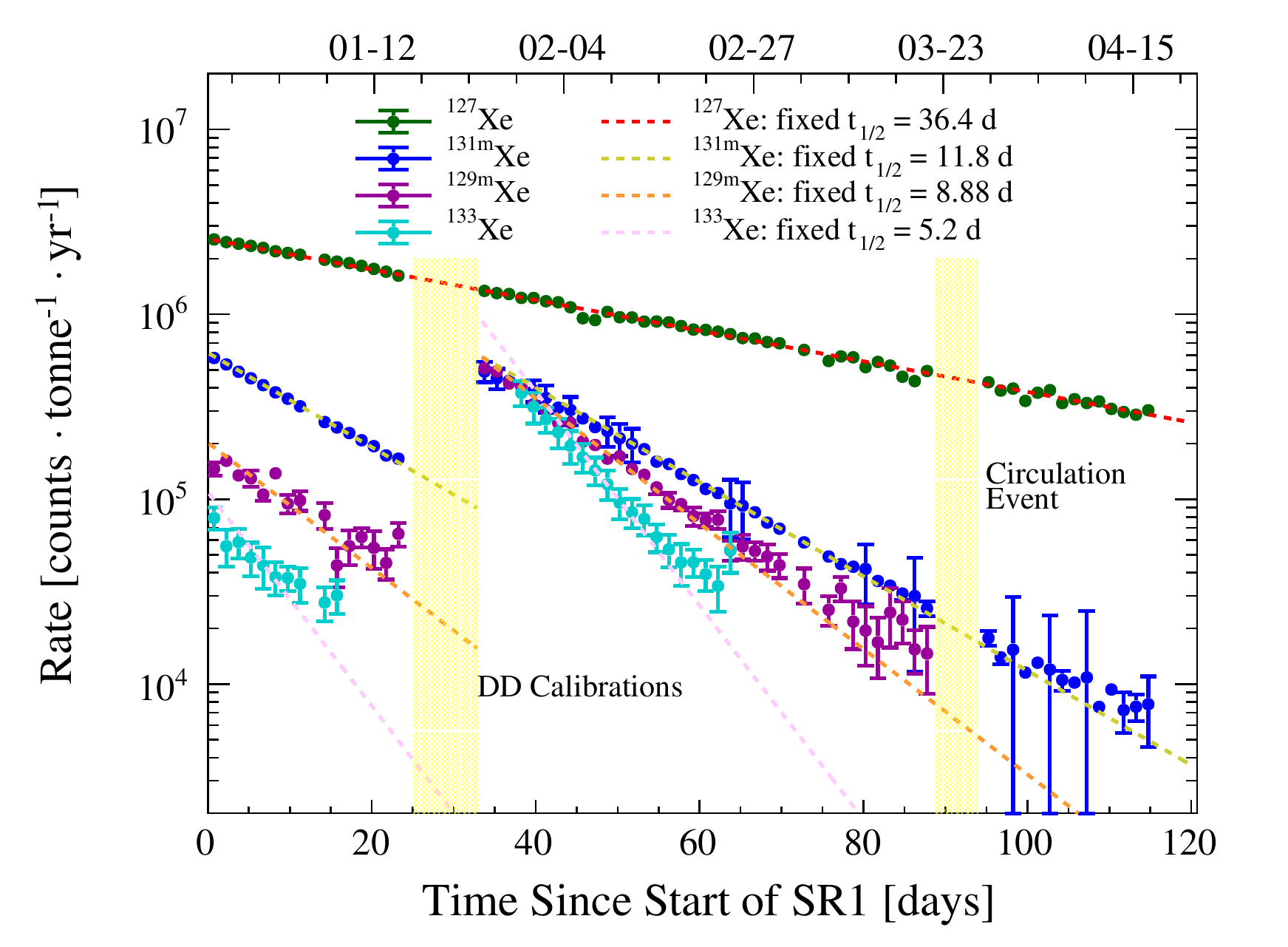}
    \caption{Rates of activated $^{127}$Xe, $^{131m}$Xe, $^{129m}$Xe, and $^{133}$Xe as a function of time in 1.5 day bins during the SR1 exposure beginning 23 December 2021. The two hiatuses are due to the mid-SR1 DD calibration, which resulted in additional neutron activation of the active xenon, and a circulation interruption that led to a decrease in purity, which affected energy reconstruction. Exponential trends for the measured half-lives are superposed.}
    \label{fig:ActivationRatesInTime}
\end{figure}

To understand the rates of the decays of $^{127}$Xe, $^{129m}$Xe and $^{131m}$Xe, and their time evolution over the SR1 exposure, their mono-energetic peaks were analyzed in the SR1 FV with fits to the combined energy spectra constructed from events labelled as either SS or MS. This dual classification selection was needed since in the case of $^{127}$Xe, which decays via electron capture, the de-excitation gamma ray of the daughter $^{127}$I can scatter sufficiently far from the X-ray or Auger electron cascade as to be identified as a distinct interaction site. The $^{127}$Xe 375\,keV gamma-ray interaction with 5.2\,keV L-shell and 33.2\,keV K-shell cascades was modeled with the sum of two Gaussians plus a linear background. Separately, the 164\,keV peak of $^{131m}$Xe, along with the $^{127}$Xe 203\,keV gamma-ray interaction with L-shell and K-shell cascades, were modeled with the sum of three Gaussians plus a linear background. $^{129m}$Xe manifests as a 236\,keV peak which heavily overlaps $^{127}$Xe and thus was inferred from the difference of the $^{127}$Xe peaks compared with what would have been expected, having normalized for their branching ratios. The $^{133}$Xe spectra is defined by the emission of a 346\,keV endpoint beta decay to the first excited state of $^{133}$Cs, which immediately relaxes via a 81\,keV gamma ray. The $^{133}$Xe rate in the SR1 FV was measured by counting events in a ``flat" section of the energy spectrum from 90--120\,keV and normalizing to the full spectrum.

\begin{figure*}[!ht]
    \centering
    \subfloat{\includegraphics[width=.45\textwidth]{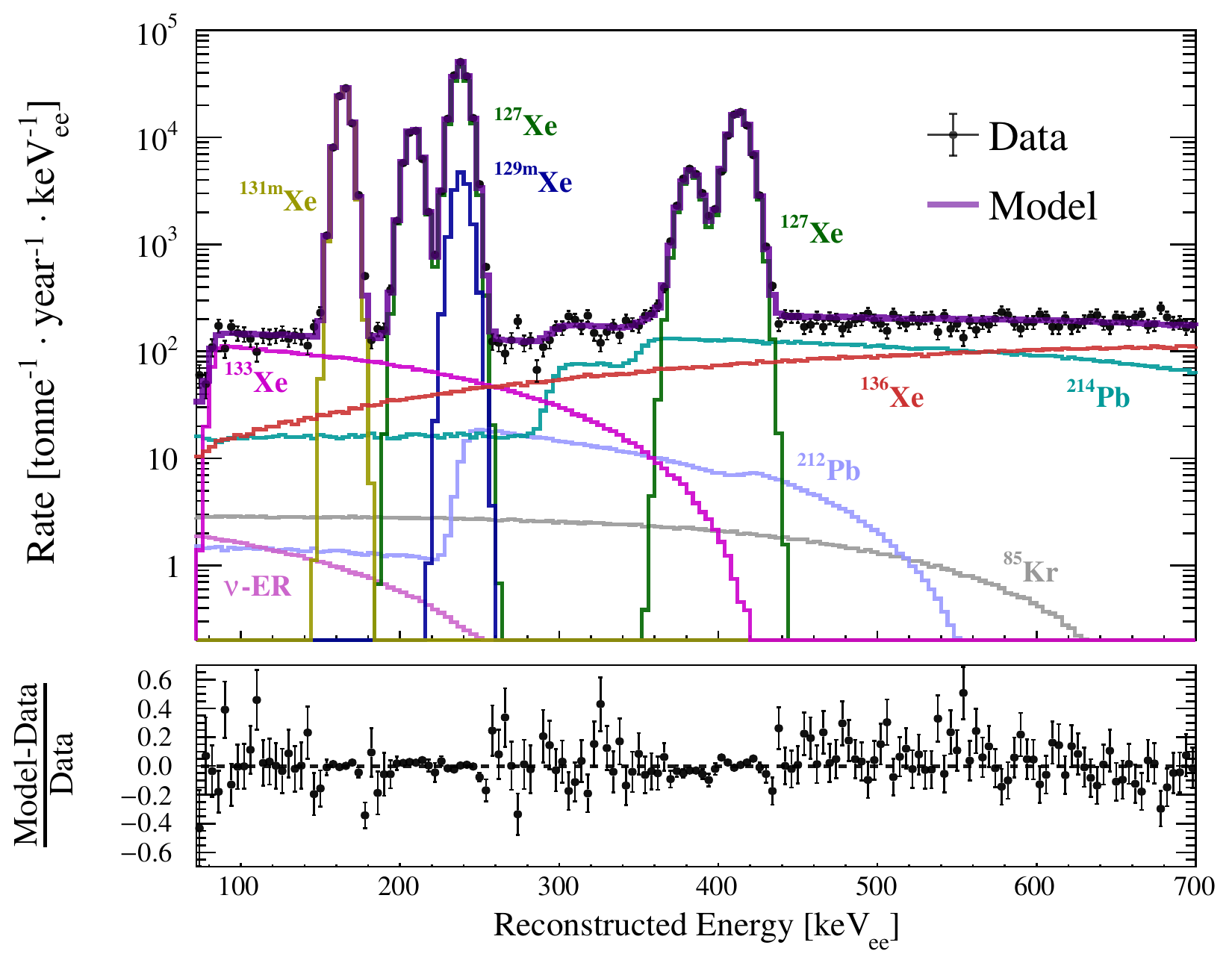}}
    \hfill
    \subfloat{\includegraphics[width=.45\textwidth]{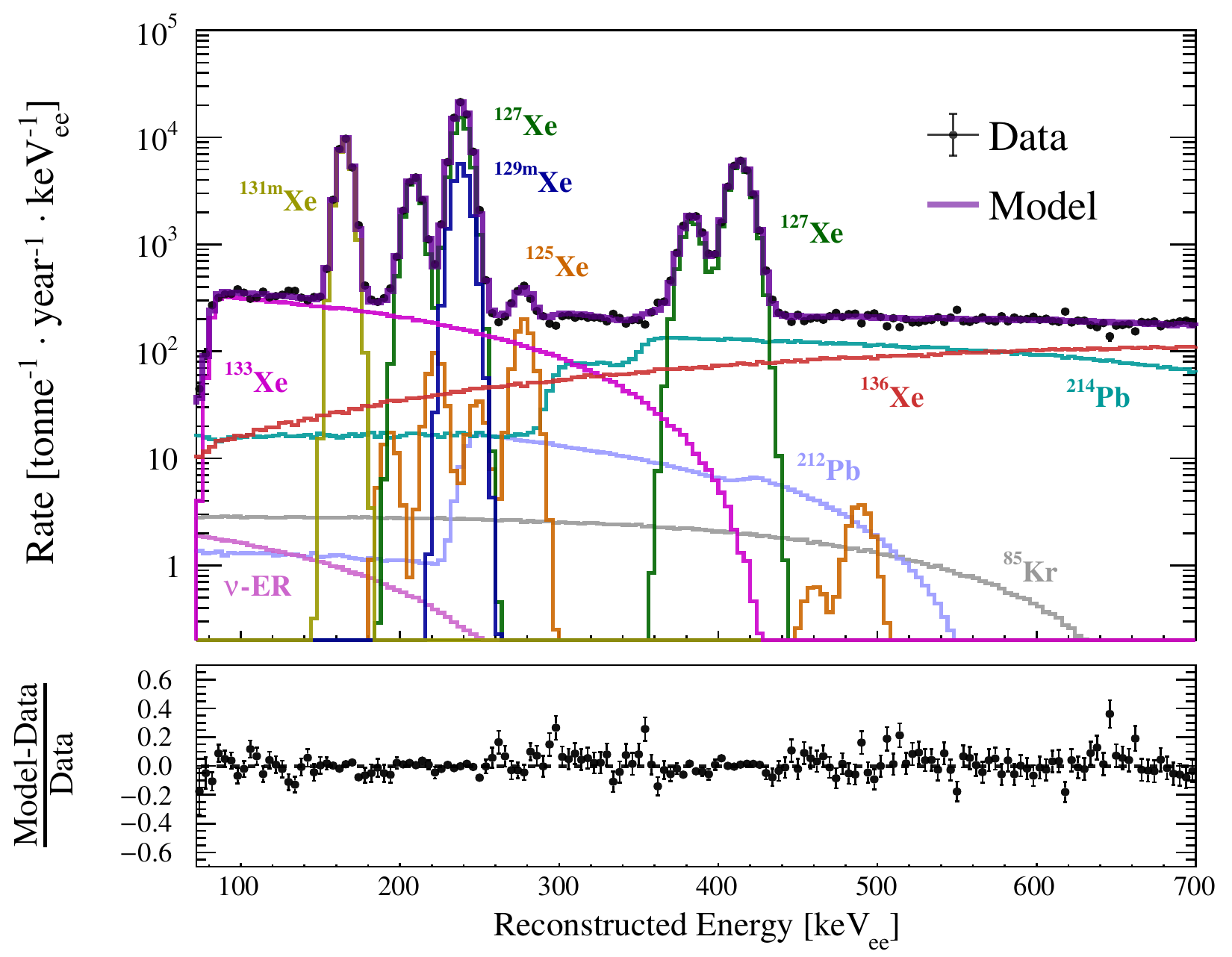}}
    \caption{Fit results for the SR1 exposure for the inner one-tonne region of the TPC. Data are shown in black and the summed background model in purple.
    \emph{Left:} Pre-DD calibrations
    \emph{Right:} Post-DD calibrations.}
    \label{fig:ModelFit_inner1T}
\end{figure*}

The observed rates in time are shown in Figure~\ref{fig:ActivationRatesInTime} alongside fixed exponential functions with the measured half-lives of 36.4, 11.8, 8.88, and 5.2 days for $^{127}$Xe, $^{131m}$Xe, $^{129m}$Xe, and $^{133}$Xe, respectively \cite{TabRad_v8,NDS-Xe131m,NDS-Xe129m}. The estimation for $^{129m}$Xe broke down towards the end of SR1 when the number of expected $^{129m}$Xe events was comparable to the statistical uncertainties in the $^{127}$Xe peak integral. Similarly, the counting of $^{133}$Xe events became unreliable when its rate was comparable to that of $^{214}$Pb and $^{136}$Xe (see Figure~\ref{fig:ModelFit_inner1T} for reference), and therefore $^{133}$Xe data below a rate of $2\times10^{4}$ counts/tonne/yr were excluded. As $^{133}$Xe was almost entirely repopulated by neutron activation from the mid-SR1 DD calibration, five days' worth of data points immediately following this calibration were also removed to account for the time needed for $^{133}$Xe to uniformly mix within the fiducial volume. This was a conservative approach given the xenon mixing estimate of around three days from observations of how injected calibrations sources, such as $^{83m}$Kr, disperse in our detector. Separate publications are being prepared on our xenon circulation and calibration systems \cite{LZ-circulation, LZ-calibration}.

The steps up in rate for $^{131m}$Xe, $^{129m}$Xe, and $^{133}$Xe after the DD calibration were expected, given the production cross-section and abundances of the isotopes. The production rate of $^{127}$Xe from neutron activation by the DD calibration was estimated to be an order of magnitude smaller than that of the metastable states, and thus any rate increase is imperceptible given its high extant concentration \cite{Naylor:2022,Taylor:2022}. This implies that the $^{127}$Xe rate can be safely extrapolated backwards in time and assumed to be of entirely cosmogenic origin: the activity at the start of SR1 was calculated as 80.4\,$\pm$\,\SI{9.0}{\micro\becquerel/\kilogram}. A $^{127}$Xe activity of \SI{76.9}{\micro\becquerel/\kilogram} at the start of SR1 was inferred from a measurement of the $^{127}$Xe rate in LUX by scaling the equilibrium value of $^{127}$Xe according to the duration and change in cosmic ray flux during surface transportation and underground storage. The details of this process are explained in Ref.~\cite{LUX-BG}. A separate calculation of cosmogenic activation with the ACTIVIA simulation package \cite{BACK2008286} yields a starting activity of \SI{52.1}{\micro\becquerel/\kilogram}. Disagreements between ACTIVIA and experimental results for $^{127}$Xe have been previously reported in Ref.~\cite{Baudis-Activation}.

\subsection{The Beta Background Region (80--700\,keV)}
\label{sec:BetaFit}

\begin{table}[!b]
    \renewcommand{\arraystretch}{1.2}
    \centering
    \caption{Inner one-tonne volume fit results for SR1, determined separately for two periods, one before and one after the DD-calibration. The errors shown in the table are the statistical errors reported from the minimization. Reported rates are those extrapolated to the start of SR1 (23 December 2021). A 10\% systematic in addition to these values can be assumed to account for uncertainties associated with the exposure estimation and event reconstruction.}
    \begin{tabular}{l c c c c }
        \hline  \hline
        Component & Half-life & Pre-DD Fit & Post-DD Fit \\
         & [days] & [\SI{}{\micro\becquerel/\kilogram}] &  [\SI{}{\micro\becquerel/\kilogram}]  \\ 
        \hline 
        $^{127}$Xe & 36.4 & 92.88 $\pm$ 0.38 & 89.65 $\pm$ 0.48 \\
        $^{131m}$Xe & 11.8 & 18.87 $\pm$ 0.13 & 108.11 $\pm$ 0.74 \\
        $^{129m}$Xe & 8.9 & 4.91 $\pm$ 0.23 & 193.04 $\pm$ 6.93 \\
        $^{133}$Xe & 5.2 & 2.01 $\pm$ 0.11 & 1467.15 $\pm$ 22.21 \\ 
        $^{125}$Xe\footnotemark[1] & 0.7 & -  & 26.70 $\pm$ 1.74 \\
        $^{214}$Pb & - & 3.05 $\pm$ 0.12 & 3.10 $\pm$ 0.10 \\ 
        $^{212}$Pb & - & 0.13 $\pm$ 0.01 & 0.11 $\pm$ 0.01 \\ 
        $^{136}$Xe & - & 3.89 $\pm$ 0.18 & 3.96 $\pm$ 0.17 \\ 
        $^{85}$Kr & - & (4.21 $\pm$ 0.42)$\cdot$10$^{-2}$ & (4.18 $\pm$ 0.42)$\cdot$10$^{-2}$ \\ 
        \hline \hline
     \end{tabular}
     \footnotetext[1]{Note that $^{125}$Xe has a 16.9 hour half-life and is only measurable in the post-DD period \cite{NDS-Xe125}. With little time to homogenize, its distribution was seen to be highly non-uniform following DD calibrations, constrained to the upper third of the TPC, therefore its one-tonne estimate is not representative.}
     \label{tab:InnerTonneFits}
\end{table}

The SR1 data were examined in a central one-tonne region of the TPC in order to directly inform the rates of sources internal to the xenon. This study aimed to constrain beta radiation sources that could contribute to the WIMP search backgrounds, as well as reaffirm the xenon activation product observations from Section~\ref{sec:XeActivation}. The central one-tonne sub-volume was defined as a right cylinder of $R\leq45$\,cm and 35\,$\leq$\,$Z$\,$\leq95$\,cm, where the assumption was made that rates measured in this region were extendable to the entire TPC. The exceptions to this were $^{214}$Pb and $^{212}$Pb, which are known to be inhomogeneously distributed (Section~\ref{sec:RadonAlphas}), and the short-lived DD activation product $^{125}$Xe, which did not persist long enough to mix into the bulk xenon. Energy spectra of SS events in this volume were analyzed within the 80 to 700\,keV region. The upper energy bound, along with the volume definition, ensured gamma-ray contributions were minimal, whereas the lower energy bound was chosen to avoid the $^{124}$Xe $2\nu$DEC and WIMP effective field theory (EFT) search regions, where analyses are in preparation \cite{LZ-Xe124, LZ-EFT}. 

Two separate fits were performed to the SR1 data split by the mid-SR1 DD calibrations, using simulated energy spectra for all components other than for $^{127}$Xe and $^{131m}$Xe, where Gaussian functions were adopted. A background-only version of the likelihood in Equation~\ref{eq:likelihood} (Section~\ref{sec:PLR}) was used, and $^{125}$Xe, which appears as a new contribution following the DD calibration, was unconstrained in the post-DD version of the fit. The fits in the two time periods are shown in Figure~\ref{fig:ModelFit_inner1T}. The goodness-of-fit was assessed using the $\chi^2_{\lambda,p}$ prescription in Ref.~\cite{Baker:1983tu}. The number of degrees of freedom (NDF) was defined as the number of data points minus the number of fit nuisance parameters. The $\chi^2/\mbox{NDF}$ values for the pre- and post-DD fits were found to be 216.72/136 and 221.54/135, respectively. 

The results of the fits, and their statistical errors, can be found in Table~\ref{tab:InnerTonneFits}, where the rates have been extrapolated to the start of the SR1 exposure to allow for a more direct comparison. This extrapolation was performed with consideration of the lengths of the pre- and post-DD exposures, given the fit results yield exposure-averaged activities. The effect of neutron activation from the DD calibration on various xenon isotopes can then be more clearly discerned than from the illustrated fits. For example, despite being neutron activated, the $^{131m}$Xe population looks diminished in the post-DD panel of Figure~\ref{fig:ModelFit_inner1T} compared to the pre-DD case as the peak represents a much longer, later exposure, and thus its live time-averaged rate is lower. The fit results were found to be consistent in their reported activity of $^{214}$Pb, despite the enhancement of shorter-lived activation products following the DD calibration.

The rates of $^{212}$Pb, $^{85}$Kr, and solar neutrino scatters were not further constrained from these fits, given they were sub-dominant components in this energy space. The $^{214}$Pb and $^{136}$Xe two-neutrino double-beta (2$\nu\beta\beta$) decay contributions were highly anti-correlated. However, the tight constraints placed on the $^{136}$Xe rate from the uncertainty on the measured half-life reported in \cite{PhysRevC.89.015502} in turn limited the $^{214}$Pb result. The determined $^{214}$Pb rate was further used for analysis of its contribution to the WIMP search, with consideration of its position distribution and thus how this one-tonne estimation should be scaled for the full TPC (Section~\ref{sec:Pb214}).

\subsection{The High-Energy Gamma Region (700--3000\,keV)}
\label{sec:HEgamma}

Gamma rays from the decays of naturally occurring radioisotopes in the cavern rock and detector components, as well as those from anthropogenic radionuclide decays within the LZ assembly, can reach the TPC and interact in the xenon. $^{238}$U and $^{232}$Th, and their progeny, $^{40}$K and $^{60}$Co are the most prevalent; screening results for these decay chains were used as the basis of starting values for the fits of these sources, as outlined in this section. The contribution of gamma rays from the cavern walls was separately informed through a dedicated analysis undertaken during early detector commissioning, when it was expected to be the dominant background. These examinations did not include consideration of ($\alpha$,$\gamma$) or ($\alpha$,n) interactions that could result in high-energy gamma rays, more energetic than the 2.6\,MeV $^{208}$Tl gamma ray that is encompassed by the upper energy limit of these studies.

\begin{figure}[!ht]
    \centering
    \includegraphics[width=\columnwidth]{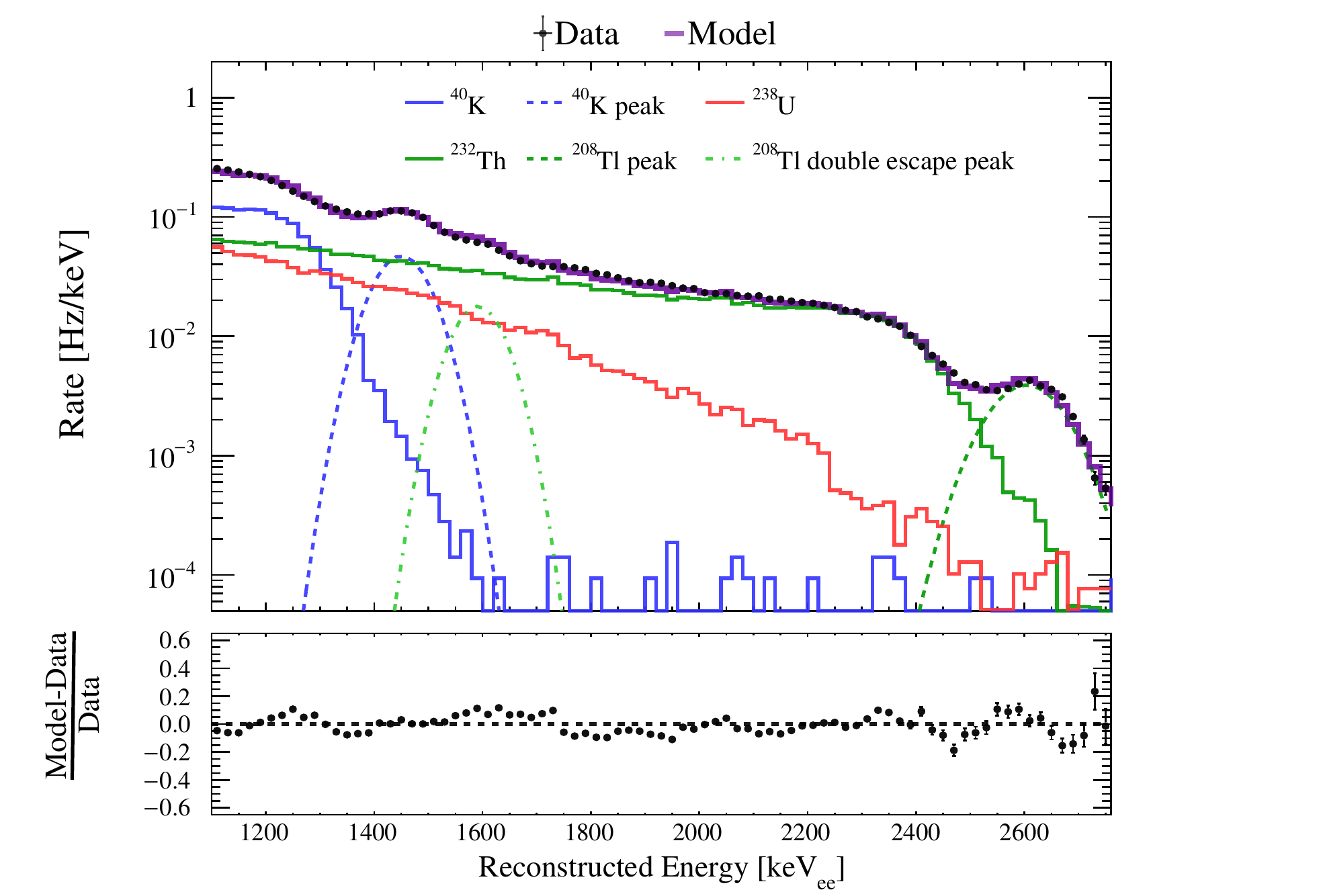}
    \caption{Fit of the cavern gamma spectra to data taken during technical commissioning when the
    TPC was filled with gaseous xenon and the water tanks and OD were empty. The reconstructed energy scale is based on the observed S1 pulses, using a linear energy calibration derived with the gamma-ray photopeak signals.}
    \label{fig:cavernGammaFit}
\end{figure}

\subsubsection{Cavern Gamma-Ray Measurement}
\label{sec:cavernGammaRayMeasurement}
Cavern rock gamma-ray normalizations were constrained via TPC measurements taken when the xenon target was in gaseous phase and the OD and water tank were both empty, therefore providing no external shielding. The event rate in the TPC was thus dominated by the $\mathcal{O}$(10\,kHz) rate of the cavern gamma-ray energy depositions: with no drift or extraction field applied, only S1 signals were observed. To correct for light collection efficiency differences in the detector, the variation in TBA of the S1s was fitted with a third-order polynomial. The polynomial was evaluated to normalize the pulse sizes to those at 0 TBA, which approximately corresponds to the midway point between the top and bottom TPC PMT arrays. A linear energy calibration was performed by fitting the mean position in S1 space of each of the visible peaks of the cavern gamma-ray spectra. 

\begin{table}[!htp]\centering
\caption{Fitted and predicted rates of cavern wall radioactivities, with the former derived from Figure~\ref{fig:cavernGammaFit}.} 
\label{tab:cavernGammaNumbers}
\begin{tabular*}{\columnwidth}{l @{\extracolsep{\fill}} c @{\extracolsep{\fill}}c @{\extracolsep{\fill}}c}\toprule
& Predicted  & Fitted  & Ratio\\
Isotope/ &  Rate &  Rate & (Fitted/\\
Chain  & (Hz/keV) & (Hz/keV) & Predicted) \\
\hline
$^{40}$K &4.2 ± 1.1 &2.79 ± 0.40 &0.67 ± 0.20 \\
$^{238}$U &3.9 ± 2.0 &1.95 ± 0.53 &0.49 ± 0.29\\
$^{232}$Th &6.1 ± 1.4 &4.51 ± 0.43 &0.74 ± 0.18\\
\hline
\end{tabular*}
\end{table}

To provide spectra to fit to the data, BACCARAT was modified for the simulation of interactions under commissioning detector conditions. These changes can be summarized in four parts: 1) the target material of the TPC was changed to gaseous xenon (GXe) and the temperature, pressure and hence density were set to the values measured during data acquisition; 2) the reflectivity of the PTFE-lined surfaces inside the TPC were changed to be representative of a GXe-PTFE interface \cite{ClaudioPTFE}; 3) the water tank and OD were emptied; 4) a GXe NEST model was used. The activity of the $\sim$149 tonne rock shell volume was set using previously measured activities of $^{232}$Th and $^{238}$U decay chains and $^{40}$K decay \cite{LZ-CavernGammas}. \geantFour particle and optical simulations of the cavern rock gamma-ray background were then completed, following the prescription in Ref.~\cite{LZSims}. Only gamma rays with energy greater than 1\,MeV were simulated to limit the computational burden, given the prominent gamma-ray lines suitable to constrain to are above this energy.

To determine the contribution of each decay chain, the component energy spectra were fitted to the data. The simulated spectra for $^{232}$Th and $^{40}$K were split into the Compton continuums and the gamma-ray photopeaks and escape peak features, to be treated as separate components in the fit, as per the prescription followed in Ref.~\cite{LZ-CavernGammas}. Similar to that analysis, the photopeak to total event ratios were required to remain within 20\% of what they were before the simulated spectra were decomposed. The final fit is illustrated in Figure~\ref{fig:cavernGammaFit}, and a breakdown of how fitted and predicted rates for each of the decay chains compare in Table~\ref{tab:cavernGammaNumbers}. The fitted rates listed consist of the sum of the post-fit continuum and peak contributions, e.g. the $^{232}$Th total rate was constructed from summing the Compton continuum, the $^{208}$Tl photopeak and the $^{208}$Tl double escape peak results. The ratio of fitted and predicted rates in the case of all three spectra considered are approximately the same.

\begin{figure}[!ht]
    \centering
    \includegraphics[width=0.9\columnwidth]{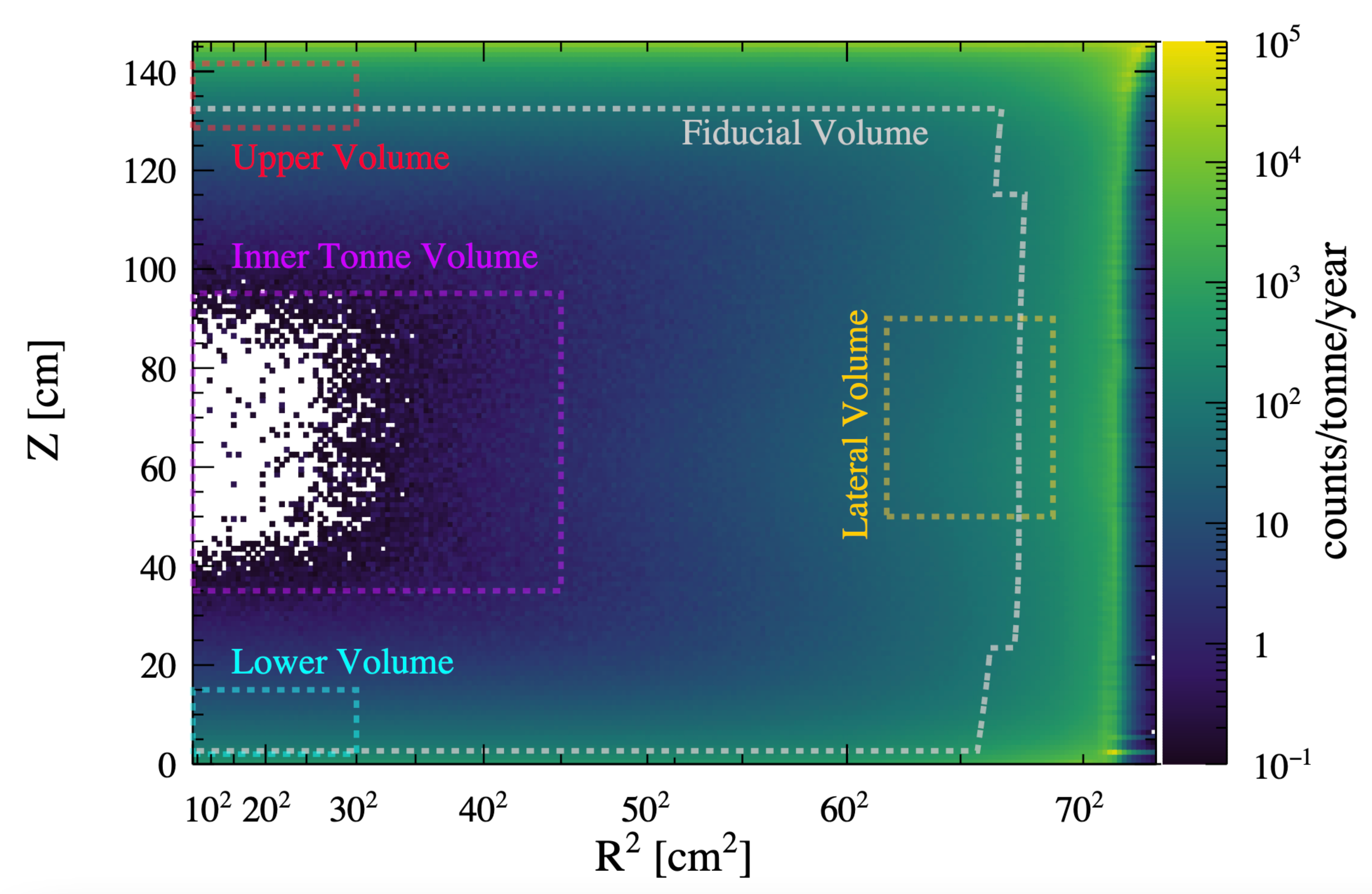}
    \caption{The simulated position distribution of all single scatter ER events from detector components and cavern gamma rays. Overlaid are the three sub-volumes (upper, lateral, lower) and the SR1 FV in which fits were performed. The inner one-tonne volume, used for the fitting of internal sources in Section~\ref{sec:BetaFit}, is also shown.}
    \label{fig:SS_DetER_FitVolumes}
\end{figure}

\subsubsection{SR1 Gamma-Ray Background Fitting}
\label{sec:SR1GammaRays}
Assessing the gamma-ray background in SR1 was complicated by the large number of possible contributions to consider, their various geometric locations, and the position dependence of the signals across the TPC. Within the contributions involving the same decay chain, variation was seen in the simulation outputs in the relative peak heights of gamma rays within their energy spectra due to the differing trajectories involved to reach the xenon target. To simplify the problem, sources were grouped together as single components to be considered for the fitting if they were of the same radiogenic origin and if their spectra were judged to be similar enough in shape, as would often be the case for isotope-location pairs originating in close physical proximity to each other. The metric for similarity was defined by a Kolmogorov-Smirnov (KS) test statistic \cite{KS_test_1, KS_test_2}, with any sources below a threshold of 0.1 grouped together. As an example, $^{40}$K sources from the top TPC array PMTs, nearby thermometers and position sensors, and the titanium plate that they were all affixed to were combined together in a single grouping.

The fitting was also broken down into separate sub-volumes, with the idea of limiting the number of prominent components to be considered in one go, before a final fit using their combined inferences was attempted in the SR1 FV. Within each sub-volume, components that did not satisfy the spectral similarity requirement and contributed fewer than 10\% of events for a given isotope were deemed subdominant, and hence were fixed in their respective fits. Three sub-volumes were defined as illustrated in Figure~\ref{fig:SS_DetER_FitVolumes}: a lateral volume to capture contributions from the TPC and cryostat walls, and a lower and an upper volume to constrain the various components present in the vicinity of the respective TPC PMT arrays. In all, these considerations were successful in reducing and compartmentalizing the 1141 simulated detector volume-decay chain pairs such that no more than a dozen parameters or groupings were fitted at any one time in a given sub-volume fit.

Fits were conducted in sequence beginning with the lateral and ending in the upper volume, using the $^{222}$Rn alpha-based corrections for the S1s and S2s used in reconstructing energy, and a background-only version of the likelihood in Equation~\ref{eq:likelihood} (Section~\ref{sec:PLR}). The considered energy range spanned from 1\,MeV to 2.7\,MeV in order to eliminate contributions from sources internal to the xenon aside from $^{136}$Xe 2$\nu\beta\beta$ decay, and to capture the highest common gamma-ray photopeak, the 2.6\,MeV peak from $^{208}$Tl. Resulting fit parameters from a given sub-volume were used as constraints for those components in successive fits. In the vast majority of cases, the exclusivity of the sub-volumes implies that the component was best fitted in a single region. This was typically the case as the components normally included sources that originated from similar locations, leading to strong position dependence of their events in the active xenon. The exceptions were the cavern gamma rays and sources from the cryostat, which irradiate a wide region of the TPC, and thus whose normalizations were seen to update from fit to fit. 

The results of these fits were used in a final SR1 FV fit (Section~\ref{sec:SR1FVFit}), in which the most substantial conclusions could be drawn about the gamma-ray background contributions.

\subsection{The SR1 FV Fit (80--2700\,keV)}
\label{sec:SR1FVFit}
Parameters produced by the series of sub-volume fits in Section~\ref{sec:SR1GammaRays} were consolidated in a final fit within the SR1 FV. The SR1 FV was selected, as opposed to the full TPC, for primarily two reasons: consistency with the SR1 WIMP search, which set a target mass in excess of what might be used in subsequent higher-energy physics searches, and in an effort to minimize unavoidable position effects at the edges of the detector, such as charge loss due to field non-uniformities. To achieve a more comprehensive picture of our background contributions, the lower energy bound of the fit was lowered to 80\,keV to encompass the rising edge of the detector components and the internal sources, whose rates were fixed to the values previously obtained by the inner one-tonne fit in Section~\ref{sec:BetaFit}. For the same reason as for that analysis, the fit was not extended below 80\,keV so as not to impinge on the energy range and expose data being considered for EFT dark matter and $^{124}$Xe $2\nu$DEC analyses. 

All simulated components were collected into categories of side, top and bottom, based on their relative event rates in each of the three sub-volumes. This led to 18 groupings: side, top and bottom for each of $^{60}$Co, $^{40}$K, $^{232}$Th-early, $^{232}$Th-late, $^{238}$U-early, $^{238}$U-late. The results from the sub-volume fits were worked in as follows: if the component had been previously constrained by a sub-volume fit, its starting normalization was reweighted according to that outcome, before it was combined with other components. The uncertainties on all components in a group, from the sub-volume fit if appropriate, else from assay measurements \cite{LZ-Cleanliness}, were combined to be used as a single Gaussian constraint for that group in the fit.

\begin{figure*}[!t]
    \centering
    \includegraphics[width=\linewidth]{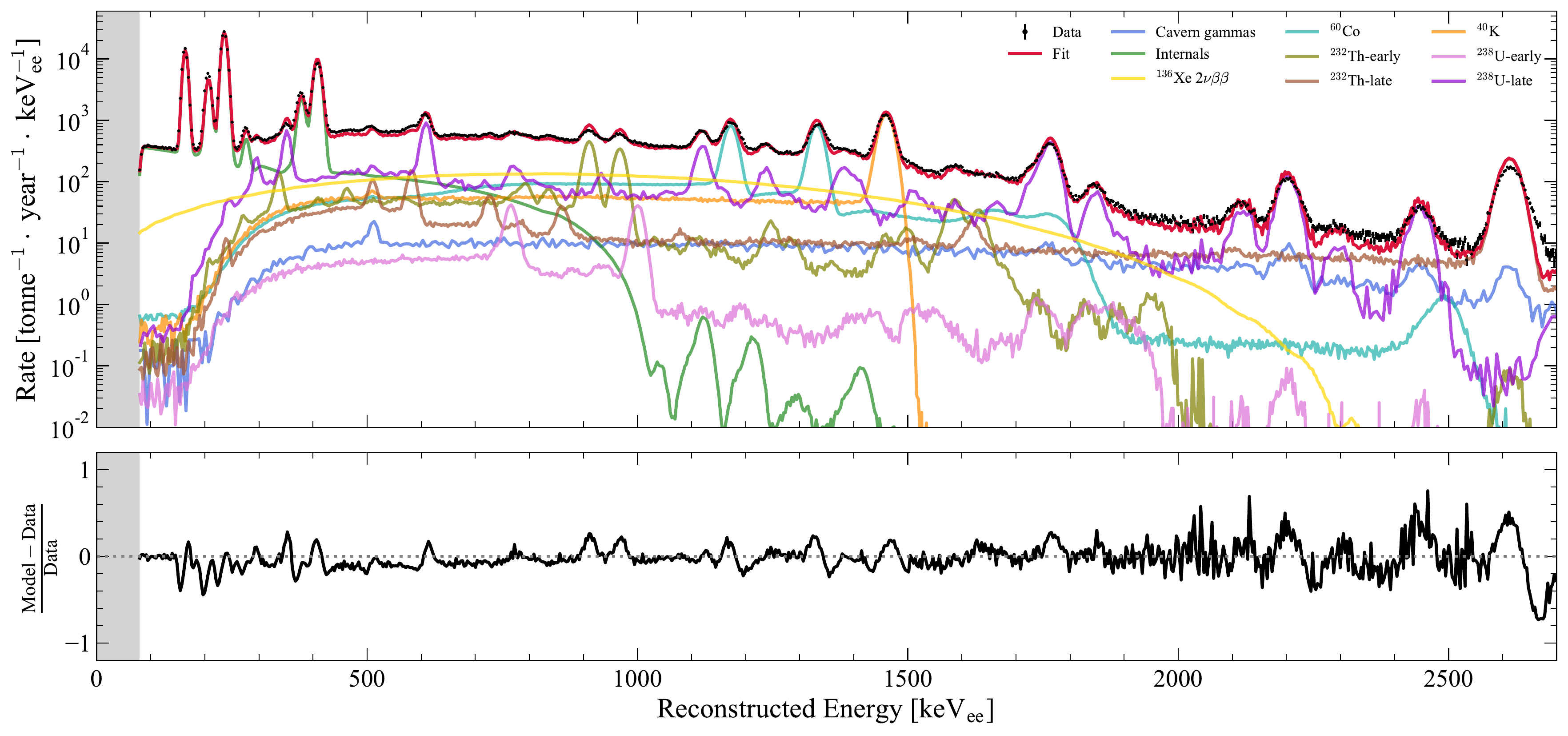}
    \caption{Fitted detector component spectra in the SR1 FV, following the prescribed sequence of sub-volume fits. The fit results are consistent with an average residual of approximately zero, with fluctuations at certain peaks arising from imperfections in energy resolution matching between data and simulations. The contribution from internals is fixed here according to the outputs of the central sub-volume fit. Gray shading is used to obscure data $<80$\,keV$_\text{ee}$ to avoid inferences in the regions of interest for future EFT and $^{124}$Xe $2\nu$DEC searches.}
    \label{fig:gamma_fit_fv}
\end{figure*}

The results of the final SR1 FV fit are illustrated in Figure~\ref{fig:gamma_fit_fv}, showing the total fitted outcomes for each isotope or decay chain. Table~\ref{tab:GammaFits} reports the total assay-estimated material activities for each of the aforementioned 18 groupings, under the ``Screening Estimate" column, and their fitted equivalents. To ensure the comparisons are meaningful, only values for components that were thought to be reasonably constrained by the fit were accumulated. The two criteria to assess this were that $>1$\% of the decays of that component should result in events in the FV, and that those events should comprise $>1$\% of all FV events  

\begin{table}[!h]
    \renewcommand{\arraystretch}{1.2}
    \setlength{\tabcolsep}{6pt}
    \caption{Cumulative source activities for different groupings of sources used in the FV fit. The screening estimates were found combining the material assay information given in Ref.~\cite{LZ-Cleanliness} for the components in each grouping, whereas best fit numbers were derived from the FV fit in Figure~\ref{fig:gamma_fit_fv}. Only components that have $>$1$\%$ of their decays create events in the FV and contribute $>$1$\%$ of all estimated counts in the FV are considered in the totals.}
    \begin{center}
        \begin{tabular}{l c c c c }
        \hline  \hline
        \textbf{Isotope/} & \textbf{Region} & \textbf{Screening} & \textbf{Best fit [Bq]} \\ \textbf{Chain} 
         &  & \textbf{estimate [Bq]} & \\ 
        \hline 
        & Top & 1.13  $\pm$  0.11	& 1.05  $\pm$  0.11 \\
        $^{60}$Co & Side & 1.18  $\pm$  0.12	& 1.12  $\pm$  1.02 \\
        & Bottom & 0.81  $\pm$  0.08	& 1.53  $\pm$  0.19 \\
        & \textbf{Total} & 3.11  $\pm$  0.18 & 3.71 $\pm$ 1.04 \\ \hline
        & Top & 7.63  $\pm$  0.76	& 2.94  $\pm$  1.66 \\
        $^{40}$K & Side & 2.56  $\pm$  0.26	& 6.32  $\pm$  0.61 \\
        & Bottom & 6.54  $\pm$  0.65	& 5.58  $\pm$  2.19 \\
        & \textbf{Total} & 16.73 $\pm$ 1.04 & 14.85 $\pm$ 2.81 \\ \hline
        & Top & 0.28  $\pm$  0.03	& 0.33  $\pm$  0.29 \\
        $^{232}$Th-early & Side & 0.66  $\pm$  0.07	& 0.66  $\pm$  0.49 \\
        & Bottom & 0.22  $\pm$  0.02	& 0.23  $\pm$  0.17 \\
        & \textbf{Total} & 1.16  $\pm$  0.07 & 1.22 $\pm$ 0.59 \\ \hline
        & Top & 0.25  $\pm$  0.02	& 0.11  $\pm$  0.16 \\
        $^{232}$Th-late & Side & 1.05  $\pm$  0.10	& 2.57  $\pm$  1.75 \\
        & Bottom & 0.30  $\pm$  0.03	& 0.32  $\pm$  0.27 \\
        & \textbf{Total} & 1.59  $\pm$  0.11 & 3.00 $\pm$ 1.78 \\ \hline
        & Top & 2.37  $\pm$  0.24	& 3.70  $\pm$  1.80 \\
        $^{238}$U-early & Side & 1.99  $\pm$  0.20	& 3.92  $\pm$  1.53 \\
        & Bottom & 1.86  $\pm$  0.19	& 2.72  $\pm$  1.40 \\
        & \textbf{Total} & 6.21 $\pm$  0.36 & 10.34 $\pm$ 2.75 \\ \hline
        & Top & 0.84  $\pm$  0.08	& 0.63  $\pm$  0.30 \\
        $^{238}$U-late & Side & 0.54  $\pm$  0.05	& 3.01  $\pm$  0.61 \\
        & Bottom & 0.95  $\pm$  0.09	& 1.28  $\pm$  0.73 \\
        & \textbf{Total} & 2.32  $\pm$  0.14 & 4.92 $\pm$ 1.00 \\
        \hline \hline
     \end{tabular}
    \end{center}
     \label{tab:GammaFits}
\end{table}

In examining position distributions of events prior to fitting, evidence was seen of a potential mismodeling issue, with proportionally more events recorded towards the side of the TPC than the top of it in simulations compared to data, and vice versa. This potential exchange of activity between side and top components may be responsible for some of the differences observed in Table~\ref{tab:GammaFits}, for example in the case of $^{40}$K. The most direct proof of this in the fit results was in the incompatibility between reported scaling factors of the cavern gamma-ray contributions, single components that irradiate the full TPC, in the lateral ($3.88\pm0.80$) versus the upper sub-volumes ($0.46^{+0.73}_{-0.46}$). The value reported in the upper volume is consistent with a value of $0.65\pm0.14$ for the GXe study that can be determined from the weighted average of the entries in Table III, where the uncertainty is just due to the statistics of the measurement. For reference, the SR1 FV fit yields a scaling factor of $0.72^{+1.55}_{-0.72}$.

The fit procedure was seen to preferentially raise the activities of certain components with respect to the initial estimates from radioassay measurements. The origin of some of these rises is understood: the elevated rates of $^{60}$Co in the bottom TPC PMTs are associated with unintentional activation when a number of them were stored nearby neutron sources. The fitted activities of $^{238}$U-early are amplified across all positional groupings. However, given their small contributions, it is likely that the fit is simply not as sensitive to these components as the constraints would suggest. The total $^{238}$U-late contribution is also elevated compared to the prediction, which may be partially explained by the quality of fit around the main gamma-ray photopeaks, where energy resolution was more mismatched between simulations and data.

The $\chi^2/$NDF for the fit is 13914.20/847, which can be compared to the pre-fit version of 187741.23/847. The quality of the fit implies that it can be used to attribute the observed counts to each decay chain, a feature useful for future high-energy searches.

\section{Backgrounds Counted in the SR1 Exposure}
\label{sec:CountedTPCBackgrounds}
\subsection{Muons}
\label{sec:Muons}

Muons are readily identified in the data due to the high-energy nature of their interactions. A preliminary study selected candidate muon events in the TPC by mandating that there should be coincident signals in both the OD and Skin. Additionally, the signals in all three detectors were required to meet a threshold on either the maximum or total integrated pulse size in the event. Applying these cuts, 1061 candidate muon events were found. This corresponds to $\sim$12 muon events per day in the TPC, seen in all three detectors, assuming high muon selection efficiency and that close to all events identified would be muons. For comparison, current simulations, using the framework detailed in Ref.~\cite{LZSims}, combined with the most recent measurement of the total muon flux at SURF by the Majorana Demonstrator \cite{MajoranaMuons}, would indicate a rate of 13.4\,$\pm$\,0.4 muons per day which interact in all three detectors, where the uncertainty quoted is solely that from the measurement. Analysis is ongoing to optimize the selection criteria and quantify their efficiencies in order to determine the muon flux at the LZ location \cite{LZ-Muons}.

Muons do not themselves pose a problem for a physics search. However, atmospheric muons that traverse the rock surrounding the Davis Campus can produce energetic neutrons which impact the LZ detector. The attenuation of the cosmic ray muon flux by the rock overburden implies at least a three order of magnitude lower production rate of muon-induced neutrons than that of radiogenic neutrons in the rock \cite{CARSON2004667}. On the other hand, muon-induced neutrons have harder energy spectra up to GeV energies, and thus can penetrate the shielding to reach the TPC. An earlier simulation study found that LZ would see 1.4\,$\pm$\,0.2 events in 1000 days from this source before analysis cuts were considered \cite{LZSims}. Therefore, muon-induced neutrons are not explored further here.

\begin{figure}
    \centering
    \includegraphics[width=\columnwidth]{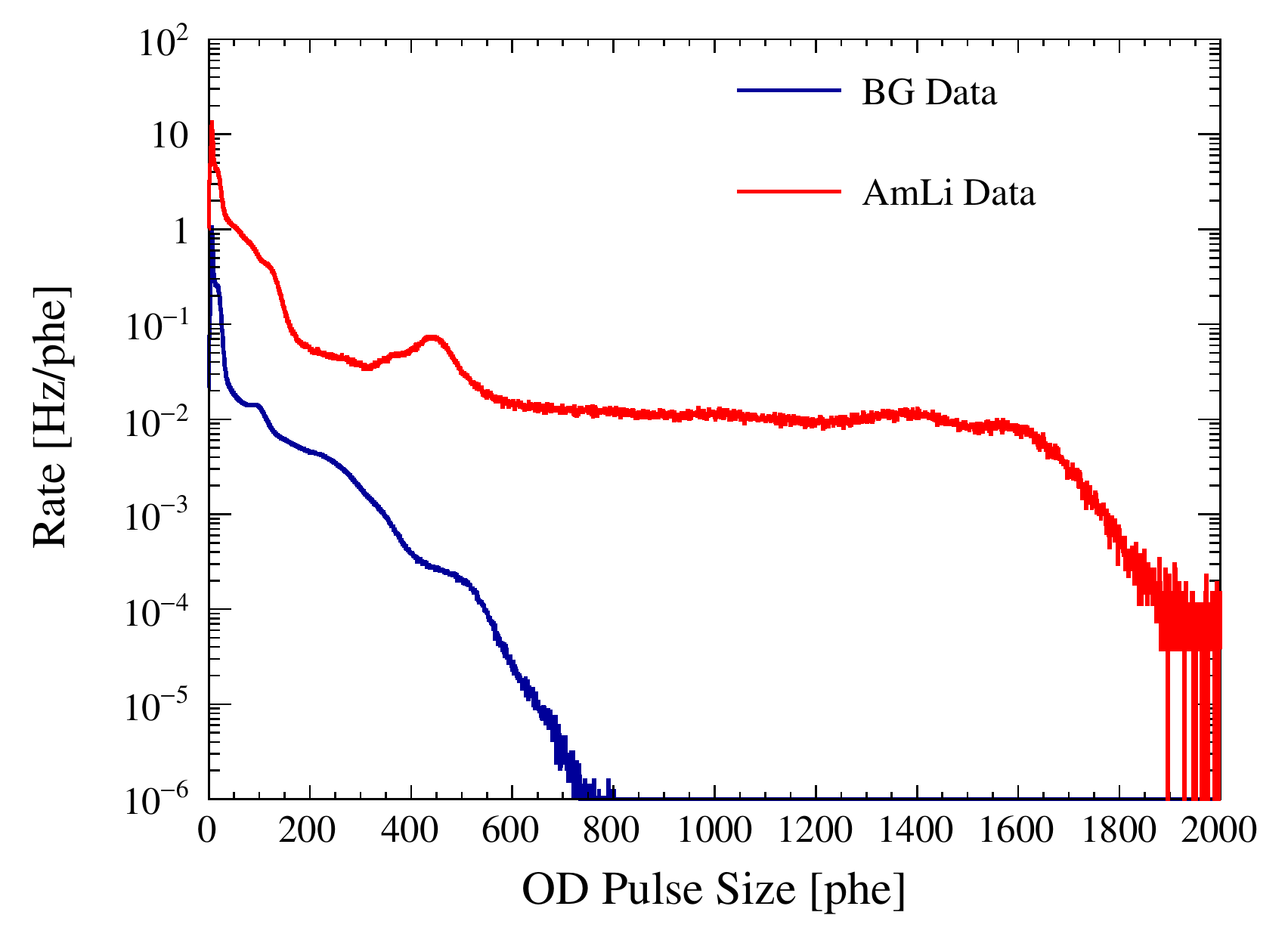}
    \caption{Total pulse size spectra in the OD for both SR1 background data and an AmLi calibration of duration 7.4 hours.}
    \label{fig:od_spectrum_amli_bg}
\end{figure}

\subsection{Neutrons}
\label{sec:NeutronCounting}
\begin{figure*}
    \centering
    \includegraphics[width=0.9\linewidth]{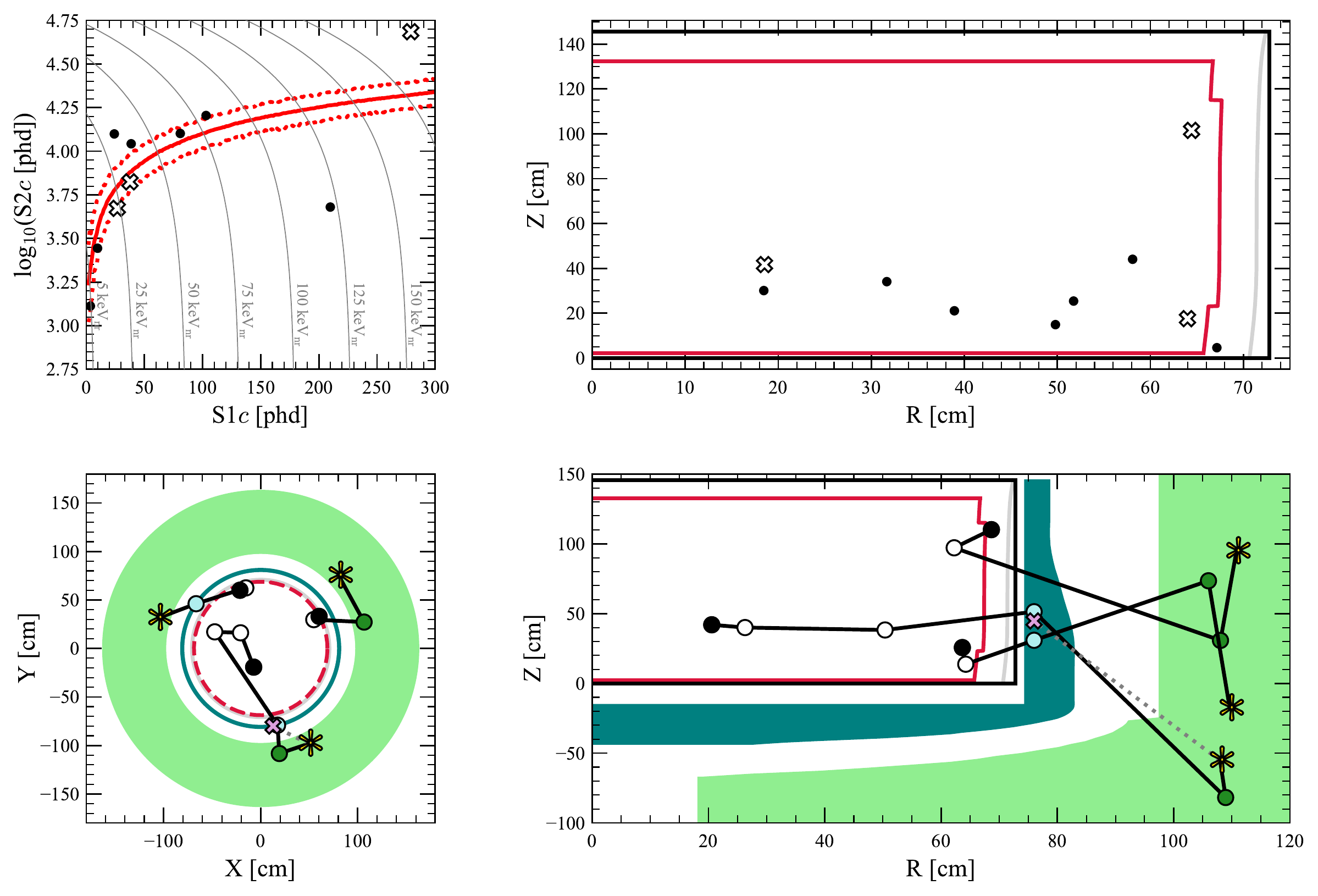}
    \caption{Locations of MS neutron events identified in the SR1 dataset, correlated across all three detectors. \textit{Top:} Distribution of the 10 identified neutron events in log$_{10}($S2$c$)-S1$c$ space overlaid with the MS NR band, as well as their averaged positions in the TPC. White crosses denote three example events displayed in detail on the second row. \textit{Bottom:} Chains of reconstructed scatters demonstrating inter-detector coincidences in tagging neutron events. Working outwards: the red outline indicates the SR1 FV; the gray curve highlights the TPC wall boundary in reconstructed space; the black box indicates the physical edges of the active xenon volume; the teal profile denotes the liquid xenon Skin; the outermost green region represents the OD acrylic tanks containing the GdLS. As the exact chronology of the event could not be determined, interactions were ordered by drift time. Black circles denote the locations of the scatters with shortest drift time in the given neutron MS chain, with empty circles showing the positions of other interactions in the TPC. Scatters in the Skin and OD are shaded in blue and green respectively. Neutron captures in the OD are marked as a $\ast$, and resulting gamma-ray splashes observed in the Skin are labeled with a pink cross. OD points are randomly assigned radial positions as XY reconstruction there is often biased towards the centre, a correction for which is under development.}
    \label{fig:NeutronDistributions}
\end{figure*}
Radiogenic neutrons can arise from $^{238}$U spontaneous fission and ($\alpha$, n) reactions on light nuclei in detector materials. To look for these neutrons in the SR1 data, two facts of their interactions were leveraged: first, neutrons with MeV energies have a mean free path of $\mathcal{O}$(10\,cm) in liquid xenon, and thus will typically scatter multiple times, with distinctly separated interaction sites; second, the majority of these neutrons are expected to scatter out of the TPC and be captured and detected in the OD. An effective search strategy was therefore to limit the pre-selection of events to those that were classified as MS and with correlated signals in the OD.

The signal spectrum in the OD was dominated by the Davis Cavern gamma-ray flux. A total pulse rate of 43\,Hz above 37.6\,phe ($\sim$200\,keV) was observed throughout the SR1 exposure, consistent within uncertainties with the sum of simulations based on cavern flux measurements performed with a NaI detector \cite{LZ-CavernGammas}, and radioassay-normalized simulations of GdLS internal backgrounds \cite{HASELSCHWARDT2019148}. Figure~\ref{fig:od_spectrum_amli_bg} illustrates this spectrum, contrasted with an example spectrum from an AmLi neutron calibration. To enhance the probability of seeing neutrons over other backgrounds, a cut was made to select only events with OD pulse sizes greater than 400 photoelectrons [phe]. This made use of two features particular to the neutron spectrum resulting from neutron captures in the GdLS: the 450\,phe hydrogen capture peak, corresponding to the emission of a single 2.2\,MeV gamma ray, and the continuum followed by an end point around 2000\,phe that is attributable to captures on Gd. The continuum is observed as, due to the thickness of the OD ($\sim$60\,cm), not all the energy from the 7.9\,MeV cascade of gamma rays released following capture on $^{155}$Gd or the 8.5\,MeV from capture on $^{157}$Gd is fully contained. Often, a few of these gamma rays will travel back into the Skin and TPC, causing responses in these detectors simultaneous with the OD capture signal.

A further criterion was applied on the time delay between the TPC interaction (the S1 observation) and the OD capture. When entering the GdLS, neutrons are usually well above thermal energies. Before capture becomes likely, they must thermalize through collisions with hydrogen in the GdLS. The time constant to capture in a pure GdLS volume, with 0.1\% Gd concentration, is \SI{30}{\us}. However, as the OD is finite in dimension, a small fraction of neutrons can find themselves in the acrylic of the OD tanks and the water surrounding the vessels and take much longer to capture. Simulations have shown that the time-to-capture distribution contains a tail of up to a millisecond \cite{Shaw:2016}. For the WIMP analysis, where it was important to ensure a high neutron veto efficiency, a cut of $<$\,\SI{1200}{\us} between the TPC S1 and OD signal was used to reject events. For the neutron search detailed here, a $<$\,\SI{400}{\us} time separation selection was adopted to encompass the vast majority of possible timings and avoid an influx of events with accidental coincidences between TPC and OD.

The TPC signal region of interest for this search was restricted to $2.5<\text{log}_{10}(\text{S2}c)<5.5$ and S1$c<500$\,phd. Neutrons interact with $\mathcal{O}$(keV) energy depositions, which will populate the lower end of this range, but the extension to 500\,phd covers the possibility of decays of the short-lived states of $^{129}$Xe (39.6\,keV) and $^{131}$Xe (80.2\,keV) stimulated by inelastic scattering. In addition, a subset of the S1 and S2 shape and parameter cuts developed in Section~\ref{sec:Accidentals} and relevant to MS events was applied to eliminate obvious accidental contributions.

After all selection criteria and handscanning to confirm event classification, ten events were found and are illustrated in Figure~\ref{fig:NeutronDistributions}, demonstrating the excellent position reconstruction in the TPC and the correlations between all three detectors. Figure~\ref{fig:NeutronDistributions} also shows where events are in log$_{10}($S2$c$)-S1$c$ space with respect to the MS NR band, which was determined with our calibration-tuned simulations, using the simulation-derived average multiplicity of scatters for neutron events. One event is seen clearly above the MS NR band at S1$c$ of 279\,phd, and is ascribed as a $^{129}$Xe inelastic scatter event. Another is seen well below at S1$c$ of 210\,phd and is a candidate “neutron-X” event, wherein the neutron, and likely the coincident gamma rays associated with the original reaction, has scattered in the S2-insensitive region below the cathode, enhancing the S1 signal relative to the S2s. The OD neutron veto efficiency at a threshold of 400\,phe and a TPC-OD delay time of $<$\,\SI{400}{\us} was established with AmLi neutron calibration analysis at $49\pm3$\%. Therefore, the total number of MS neutron interactions this observation implies were present in SR1, agnostic to veto response, is $\sim$20 events. The observed weak bias of the events towards the bottom of the TPC is being investigated.

\section{Background Impact on the SR1 WIMP Search}
\label{sec:WIMPBackgrounds}
To perform the WIMP search (WS), a number of selections were made on the SR1 data to exclude non-candidate events, as described in detail in Ref.~\cite{LZ-SR1result}. The analysis was conducted using SS-classified events, in a region of interest (ROI) of $3<\text{S1}c<80$\,phd, log$_{10}$(S2$c$) $<$\,5, and $\text{S2}_\text{raw}>600$\,phd, where `raw' specifically denotes the pulse has not been position-corrected. An additional three-fold PMT coincidence requirement was imposed on the S1s. This ROI encompasses the expected signals of WIMP interactions for masses of a few GeV/c$^2$ up to $\mathcal{O}$(TeV/c$^2$). The SS classification efficiency, as it varies with energy in the WIMP ROI, was evaluated using neutron calibration data as discussed in Ref.~\cite{LZ-SR1result}.

The SR1 data were subjected to further analysis cuts to improve the quality of the persistent data and to remove events with identifiable background features within this space. Veto anti-coincidence cuts preserved events that did not see a response in the OD greater than 37.6\,phe (200\,keV) within \SI{1200}{\us} after the TPC S1, mitigating neutrons, and those events that did not see a signal within $\pm$\,\SI{0.3}{\us} (\SI{0.5}{\us}) of the TPC S1 pulse in the OD (Skin), that could be potentially created by gamma rays associated with the TPC event. 

A fiducial volume was implemented to combat wall backgrounds that could be misreconstructed towards the centre of the TPC, and to remove external backgrounds, whose position distribution was skewed towards the edges of the detector. The FV definition was $86<\text{drift time} <\text{\SI{936.5}{\us}}$ and radius $>$\,4.0\,cm from the TPC wall in observed space, with extra cut outs of radius $>$\,5.2\,cm for drift time $<$\,\SI{200}{\us} and $>$\,5.0\,cm for drift time $>$\,\SI{800}{\us}. Being strongly background-motivated, the FV choice is discussed in Section~\ref{sec:WallFV}. 

A suite of S1 and S2 pulse-based cuts were developed to tackle pulse pathologies that typically manifest in accidental coincidence events, in which unrelated S1s and S2s would be falsely paired and classified as a SS event; these are discussed in further detail in Section~\ref{sec:Accidentals}. In addition, a series of live time vetoes was applied that discarded high pulse rate periods, wherein the probability of such fake SS events was elevated due to the increased incidence of pulses that can mimic S1s and S2s formed by the pile-up of other pulses. Combined, these vetoes result in a reduction of the live time to an effective $60\pm1$ days. 

After all cuts were applied, 335 events remained in the SR1 WS dataset on which a statistical inference was performed in log$_{10}($S2$c$)-S1$c$ space (Section~\ref{sec:PLR}). The analysis therefore required information on the expected distribution of each background in this observable space, as well as their rates. The former was obtained through simulations of each component (Section~\ref{sec:Sims}), on which cuts were applied in accordance with the data treatment. As only S1 and S2 sizes and not pulse shape information were simulated, acceptance functions were defined for pulse-based cuts through studies on calibration data and were applied to the simulation outputs. Normalizations for these probability distribution functions (PDFs) were informed through analysis of the SR1 data; many of these rely on more global observations of the backgrounds as outlined in Section~\ref{sec:TPCBackgroundFits}. These rates were characterized outside of the WIMP ROI to avoid experimenter bias. 

Aside from the wall background rate, which was factored into the FV choice in Section~\ref{sec:WallFV}, specific derivations of the normalizations are described in Sections~\ref{sec:BetaAndGamma} to \ref{sec:NeutronConstraint}, with the various categories of background covered in descending order of expected dominance in the WS background budget.

\subsection{Fiducial Volume Definition and the Wall Background}
\label{sec:WallFV}

The development of the fiducial volume cut for SR1 was highly driven by background considerations. External radiation is attenuated as it enters the TPC due to the self-shielding power of the xenon. Event rates in SR1 were therefore elevated towards the edges of the target and could be effectively mitigated through fiducialization. In the vertical direction, cuts were defined at drift times of \SI{936.5}{\us} and \SI{86}{\us}, above and below which events were excluded. These drift times correspond to heights 2.2\,cm above the cathode at the bottom of the forward field region of the TPC, and 12.8\,cm below the gate grid, respectively \cite{LZ-detector}. The lower drift time cut ensured the removal of cathodic events in the active region of the detector. The upper drift time cut was motivated by the need to remove SS events originating close to the liquid-gas interface, as well as those from the gaseous phase between the anode grid and the top PMT array that had been misreconstructed into the active volume. These cut definitions were adopted to be equivalent to those assumed for projected sensitivity studies in Ref.~\cite{LZ-sensitivity} and therefore not further optimized for SR1. 

Updated simulation-based studies using this lower drift time bound confirmed an expected zero counts in the SR1 exposure of so-called “gamma-X” events. These events are a potential WIMP background in which the gamma ray also scatters in the S2-insensitive region below the cathode, adding to the S1 signals and producing SS-classified events with more NR-like ratios between the S1 and S2. Given the centimeters of liquid xenon between the cathode and the bottom edge of the fiducial volume, only gamma rays with energies $\mathcal{O}$(100\,keV) and greater contribute to the gamma-X rate, causing this background in the $\text{S1}c<80$\,phd region to be virtually non-existent.

The more critical dimension to determine was the radial boundary because of the possibility of backgrounds physically outside the boundary being misconstructed within it, due to the resolution of the S2-determined XY position. As smaller S2s have worse position resolution, this problem was exacerbated by charge loss at the walls. Small, local field non-uniformities existing between the field cage rings that establish the TPC drift field prevent the full extraction of the charge signals near the walls, resulting in smaller amplitude S2s. 

Simulations of the drift field indicated that events up to 3\,mm away from the TPC wall could exhibit charge loss. Hence, not only events originating from the surface contamination on the PTFE could be impacted, but also those of external gamma rays stopped within this outer layer of xenon. The concern is if affected events have increased S1-S2 ratios such that they manifest in the NR band.

The position of the TPC active volume wall in observable space for SR1 was defined examining $^{83m}$Kr calibration events with a reconstructed radius $>65$\,cm. Moving radially outwards, the event distribution remains constant before gradually decreasing on approaching the detector edge, where the reconstruction uncertainty smears the reported positions. The wall position was defined
where the reconstructed R$^2$ distribution reaches half of the uniform density of the interior of the detector.

This procedure was repeated for drift time slices of width \SI{32}{\us}, and the full reconstructed wall position obtained by fitting these drift time bin values with an empirical function. The wall appears curved in reconstructed coordinates because of the curvature of the underlying drift field that governs the XY positions reported for the S2s. No azimuthal dependence was seen in the $^{83m}$Kr events, and therefore the observable wall position was defined as a function of drift time only.

 \begin{figure}[!h]
         \centering
         \includegraphics[width=0.5\textwidth]{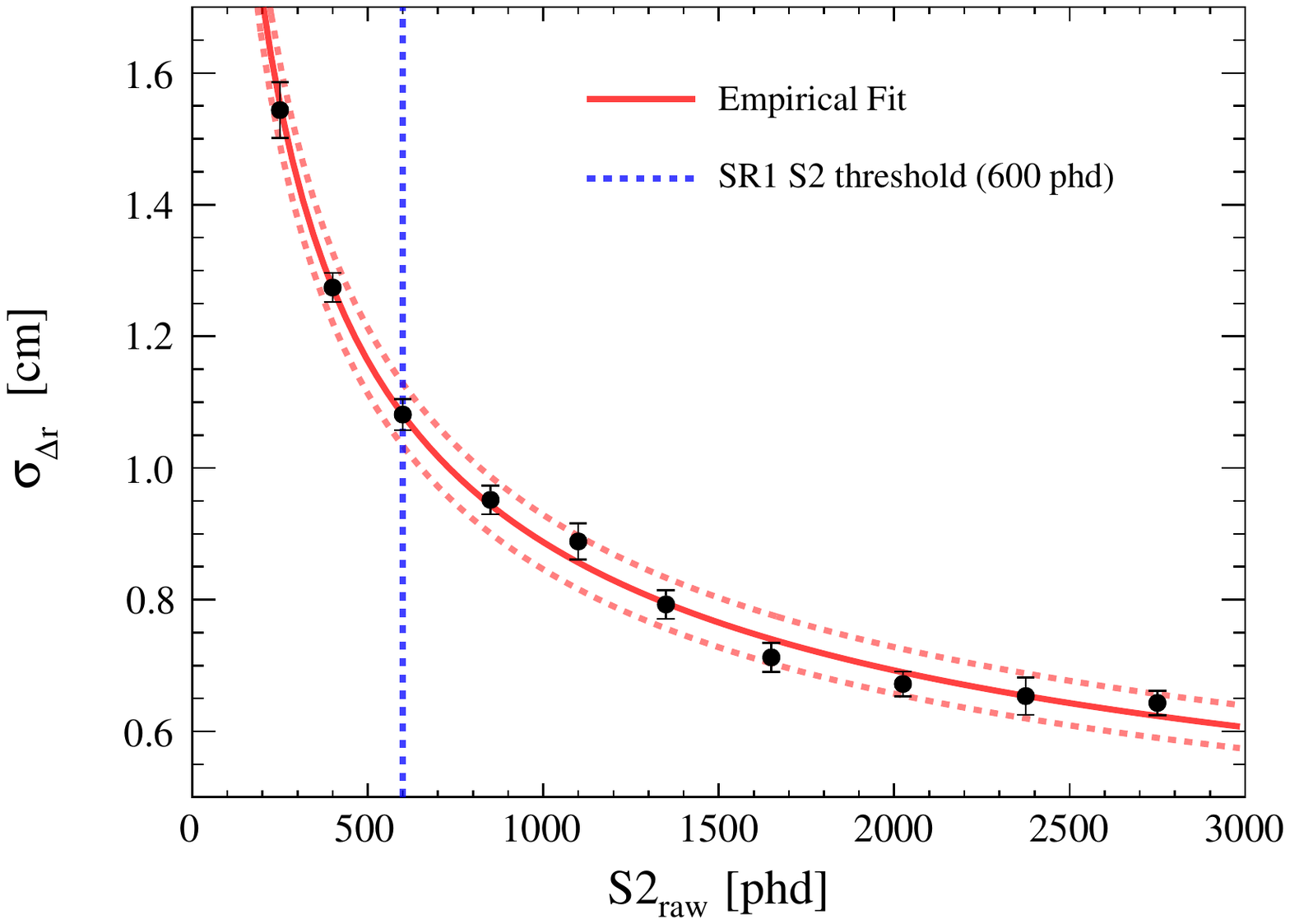}
         \caption{Dependence of the radial position resolution of the wall events on S2 pulse size. Events below the ER band and in the WS sideband region ($80<\text{S1}c<500$\,phd) were selected to determine the radial position resolution of wall events, and were observed to depend on the S2 pulse size based on the equation: $\sigma_{\Delta r}$ (S2) = $\frac{a}{\sqrt{S2}} + b$, where the fitting parameters $a$ and $b$ were $21.04\pm0.65$ and $0.22\pm0.02$, respectively. }
         \label{fig:Wall_S2position_resolution}
     \end{figure}

In establishing the S2 position resolution model, charge loss events below the ER band in the WIMP ROI adjacent region of $80<\text{S1}c<500$\,phd were considered. It was verified that their distribution was centered on the calculated wall position before their observed positions were used to evaluate the position resolution for bins of different S2 pulse sizes. The results are illustrated in Figure~\ref{fig:Wall_S2position_resolution} and follow an expected $\sfrac{1}{\sqrt{S2}}$ dependence. As the position resolution is independent of S1 pulse size, this model was adopted to investigate the wall background rate as a function of potential fiducial radius within the WIMP ROI of S1$c<80$\, phd.

The radial fiducial volume cut and the S2 threshold were simultaneously optimized, with the end goal of preserving as large a fiducial mass as possible whilst ensuring that any potential wall NR background could be safely ignored in a WIMP analysis. The estimated wall leakage was therefore desired to be $<$\,0.01 events.

The study to set the radial boundaries used events in the sideband region of $100<\text{S1}c<500$\,phd that passed all data quality cuts used for the WIMP analysis, apart from the S2 threshold. To account for the observed overdensities of events at the top and bottom of the wall, three drift time bins were considered separately in the radial optimization: $<$\,\SI{200}{\us}, 200\textendash\SI{800}{\us}, $>$\,\SI{800}{\us}. Gaussian functions were fitted to the populations in each of these bins in order to determine suitable options for these radial cuts to obtain a 0.01 wall event count. The radii selected were 5.2\,cm, 4.0\,cm and 5.0\,cm away from the wall, respectively in each of the drift time bins, with an S2$_\text{raw}$ threshold of 600\,phd. These boundaries were adopted for SR1, along with extra cuts to remove events clustered within a 6.0\,cm locus of the high-activity TPC field cage resistors. These latter cuts had minimal impact on the resultant fiducial mass, which was evaluated as $5.5\pm0.2$ tonnes.

An additional analysis, using the S2 position resolution model, was conducted to corroborate these radial cut choices, directly examining the WIMP ROI events to evaluate the expected wall event leakage. Within each of the three drift time bins, events below the NR band ($\text{S2}_\text{raw}<3000$\,phd) were further sub-divided into bins of 100\,phd. The wall events within $r_{\text{wall}} - \sigma_{\Delta r}$ were counted for each S2 bin, where $r_{\text{wall}}$ was the value obtained for the empirical function defining the TPC active volume wall and the $\sigma_{\Delta r}$ was obtained from the S2 position resolution model defined in Figure~\ref{fig:Wall_S2position_resolution}. For the SR1 radial FV boundaries, this method calculated a wall leakage of 0.015 events.

The radial fiducial cut will impact other backgrounds that were not used to motivate its choice: decays near the FV edge where only partial energy is captured in the xenon as associated gamma rays escape the TPC (“semi-naked” decays); the nuclear recoils of $^{206}$Pb ions expelled into the TPC from decays of $^{210}$Po embedded on the walls. Semi-naked decays of concern to the WS are those of $^{127}$Xe and $^{214}$Pb; the rates of these were constrained in Sections~\ref{sec:Xe127} and \ref{sec:Pb214}, respectively. $^{206}$Pb recoils fall under the selection of events discussed for charge loss and were not separately considered. Though once an important background in noble TPC-based dark matter experiments \cite{BRADLEY2015658}, given the confirmation that the $^{210}$Po PTFE surface activity met our construction requirement (Section~\ref{sec:RadonAlphas}), the expected $^{206}$Pb NR background in SR1 was $<$\,0.003 counts.

\subsection{Beta and Gamma-Ray Backgrounds}
\label{sec:BetaAndGamma}
Background contributions from beta and gamma-ray sources produce flat spectra that are essentially indistinguishable in the low-energy WIMP ROI. For the statistical analysis detailed in Section~\ref{sec:PLR}, these contributions were therefore combined into a single component. For more accurate evaluations of the fraction of each beta spectrum that is captured in the WIMP ROI, the spectral shapes of $^{214}$Pb, $^{212}$Pb, and $^{85}$Kr were updated based on the prescription in Ref.~\cite{Haselschwardt:2020iey}.

\subsubsection{$^{222}$Rn Chain ($^{214}$Pb)}
\label{sec:Pb214}
The $^{222}$Rn chain contains beta-emitting radioisotopes, which could generate low-energy ERs in the WIMP ROI. $^{214}$Bi is not a concern as it is followed in quick succession by $^{214}$Po, which has a half-life of \SI{164}{\us}. Given the 4.5\,ms event acquisition window used in SR1, the probability of the $^{214}$Bi beta signal being isolated in its own event window and being in the WIMP ROI is negligible. $^{210}$Pb has a 22-year half-life and thus will be effectively removed from the bulk before decaying. However, it can plate out on surfaces. The possibility of $^{210}$Pb and its progeny leaching off of surfaces has not been investigated here. Our measurement of the $^{210}$Po rate on the TPC wall ($2.32\pm 0.15$\,mBq, Section~\ref{sec:RadonAlphas}) is lower than that reported for LUX Run 3 ($>$\,14.3\,mBq) in Ref.~\cite{BRADLEY2015658} for a much greater surface area and target mass, indicative of a lower surface contamination rate. A first-principles modeling effort aimed at quantifying low-energy grid backgrounds revealed that the gate and cathode data were consistent with an origin primarily from the $^{210}$Pb chain, with an activity per grid area of $7.3\pm$\SI{0.4}{\micro\becquerel/\cm^2} and $4.3\pm$\SI{0.4} {\micro\becquerel/\cm^2}, respectively \cite{LZ-GridPb210}. Again, these numbers are lower than reported for LUX, where no evidence of $^{210}$Pb mobility was observed in Ref.~\cite{OliverMallory:2021}. Therefore, the only contribution of $^{222}$Rn chain decays considered relevant to the SR1 WS ROI is that of $^{214}$Pb.  

$^{214}$Pb undergoes “naked” beta decay with a branching ratio of 11.0\% \cite{ZHU20211}. This is a decay mode where no accompanying gamma rays are observed, and which therefore directly contributes to the WIMP search background. The beta decay topologies of $^{214}$Pb are difficult to identify \textit{in situ}, due to the lack of sharp spectral features in the beta decay continuum, thus constraints are derived indirectly through the alpha-emitting isotopes that precede and follow the $^{214}$Pb, and through energy-spectrum fitting. The $^{218}$Po and $^{214}$Po rate in the TPC were measured at 4.82\,$\pm$\,\SI{0.34}{\micro\becquerel/\kilogram} and 2.65\,$\pm$\,\SI{0.19}{\micro\becquerel/\kilogram} (Section~\ref{sec:RadonAlphas}), respectively, which bound the $^{214}$Pb rate. These were used to inform the fitting, which yielded a rate of 3.10\,$\pm$\,\SI{0.10}{\micro\becquerel/\kilogram} in a central, one-tonne volume (Section~\ref{sec:BetaFit}). The distribution of $^{222}$Rn was found to be inhomogeneous in the TPC due to xenon circulation patterns, and successive radioisotopes could be positively ionized, drifting downwards towards the cathode. Simulations of this effect were verified for $^{214}$Po and used to infer the position distribution of $^{214}$Pb, illustrated in Figure~\ref{fig:Radon_positions}c. Accounting for this result, the rate in the FV was evaluated at 3.26\,$\pm$\,\SI{0.09}{\micro\becquerel/\kilogram}.

“Semi-naked” events of gamma ray-accompanied decay modes, in which the gamma ray(s) is not observed in the TPC, were also considered. This estimate used the information gained from the study of this topology for $^{127}$Xe (Section~\ref{sec:Xe127}), given the similar principal gamma-ray energies of the two decays. Conservatively integrating the predicted non-naked $^{214}$Pb rate within 5\,cm of the radial FV definition, the radius from which $^{127}$Xe semi-naked events were observed, and taking into account the reported 22\% veto inefficiency for $^{127}$Xe (Section~\ref{sec:Xe127}) and the mean free paths of the gamma rays in xenon, the number of counts for SR1 is expected to be $<1$. Thus, the semi-naked topology was discounted in the estimation of final SR1 counts. The predicted WS counts for the naked decay mode-only in SR1, regarded as the total contribution of $^{214}$Pb, were found to be $164\pm35$, enfolding all sources of systematic uncertainty.

\subsubsection{$^{220}$Rn Chain ($^{212}$Pb)}
\label{sec:Pb212}
The $^{220}$Rn chain is analogous to the $^{222}$Rn chain. $^{212}$Bi is again not problematic to the WIMP search as, given the 300\,ns half-life of its daughter, $^{212}$Po, it cannot be present alone in an event. Aside from $^{212}$Pb, the only other radioisotope that beta decays is $^{208}$Tl, which does not have naked decay modes. Thus only $^{212}$Pb was considered. Concentrations of $^{220}$Rn were expected to be much lower than $^{222}$Rn as its shorter half-life impacts the amount that can emanate before it decays in the material \cite{LZ-Cleanliness}, implying lower activities of $^{212}$Pb compared to $^{214}$Pb. The suppression of the $^{220}$Rn chain was confirmed with the measured activities of $^{216}$Po and $^{212}$Po versus their $^{218}$Po and $^{214}$Po counterparts in the $^{222}$Rn chain (Table~\ref{tab:Radon_AlphaActivities}). 
Constraining the $^{212}$Pb rate was complicated by the fact that $^{216}$Po can decay upstream in the circulation system such that the sufficiently long-lived $^{212}$Pb progeny (10.6 hour half-life) can flow and mix in the TPC prior to decaying. The $^{212}$Po position distribution was seen to be similar to $^{222}$Rn, confirming similar mixing in the TPC. Assuming the EXO-200 $\beta$ positive ion fraction of $76.4\pm5.7$\% in Ref.~\cite{EXO-200chargedions} derived for the $^{214}$Pb to $^{214}$Bi decay could be applied to $^{212}$Bi, given our previous confirmation of the $\alpha$ positive ion fraction in Section~\ref{sec:RadonAlphas}, analogous flow calculations were performed to infer the $^{212}$Pb rate given the observed $^{212}$Po rate and distributions. The $^{212}$Pb rate was thus calculated to be 0.137\,$\pm$\,\SI{0.019}{\micro\becquerel/\kilogram}. Considering the naked beta decay branching ratio of 11.9\% \cite{AURANEN2020117}, and assuming semi-naked decays to be negligible, the number of estimated SR1 counts was $18\pm5$.

\subsubsection{$^{85}$Kr and $^{39}$Ar}
\label{sec:Kr85}
Natural xenon contains trace amounts of the beta-emitting radioisotopes $^{85}$Kr and $^{39}$Ar, which are uniformly dispersed throughout the liquid xenon and contribute to the total ER background. Both isotopes have concentrations that were greatly reduced via a charcoal chromatography campaign at SLAC. The concentrations of krypton and argon were measured at both SLAC and SURF using a liquid nitrogen cold trap, sampling each storage pack that was added to the xenon inventory of the experiment. The sampling results were consistent; accounting for both bottle measurements and residual xenon in the circulation system, the mean concentrations were determined to be 144\,ppq g/g $^{\text{nat}}$Kr/Xe and 890\,ppt g/g $^{\text{nat}}$Ar/Xe, with a systematic uncertainty on the sampling rate of 15$\%$. Additionally, periodic sampling during SR1 allowed for assessing the ingress rate of $^{\text{nat}}$Kr and $^{\text{nat}}$Ar, which was found to negligible for the short SR1 exposure. Given a 10.75 year (269 year) half-life and $2\times10^{-11}$ ($8\times10^{-16}$) natural abundance \cite{etde_5476105}, the $^{85}$Kr ($^{39}$Ar) rate in the TPC was determined to be 42.3\,nBq/kg (0.876\,nBq/kg). $^{39}$Ar was excluded from the final background model since its contribution was extremely subdominant ($<$1 event in SR1). 

The $^{85}$Kr rate was further validated \textit{in situ} by counting coincident beta-gamma ray decays. $^{85}$Kr undergoes naked beta decay to $^{85}$Rb with a 99.57$\%$ branching fraction. In the other 0.43$\%$ of decays, $^{85}$Kr beta decays to metastable $^{85m}$Rb, with a \SI{1.015}{\us} half-life, which subsequently relaxes to $^{85}$Rb via emission of a mono-energetic 514\,keV gamma ray. This delayed coincidence signal is unique to $^{85}$Kr and provides a distinct event topology of a small beta S1 followed by a larger gamma-ray S1. Events with at least two S1 signals and at least one S2 signal were therefore considered for this analysis. Strict analysis cuts were applied on the coincident S1 signal time separation, known beta and gamma-ray S1 sizes, and correlation of the TBA of the S1 pulses, based on the expected signal from the tuned detector simulation model. Considering the combined efficiency of these cuts, the SR1 counts were found to be $8.2\pm4.1$ events. This result is equivalent to an average concentration of $136\pm69$\,ppq g/g $^{\text{nat}}$Kr/Xe, in strong agreement with the sampling results. For the SR1 WS, the $^{85}$Kr rate derived from sampling measurements was used to determine the contribution to the WIMP ROI as $32\pm5$ counts.

\subsubsection{Detector and Cavern Gammas}
\label{sec:DetROIBkgs}
Gamma rays originating from trace amounts of $^{60}$Co, $^{40}$K, $^{238}$U, and $^{232}$Th present in the detector materials, in addition to those of $^{238}$U, $^{232}$Th and $^{40}$K from the cavern walls, contribute to the ER rate in the WIMP ROI and are collectively referred to as the Detector ER background. With effective fiducialization, the Compton plateau contribution is expected to be small. The extensive simulations undertaken for each component of the Detector ER background were scaled with the results from fits at higher energies, in which these contributions are dominant (Section~\ref{sec:HEgamma}). For the estimation of counts, all cuts were applied to the simulation outputs, including veto rejection. This is the only source for which veto cuts were applied directly from simulations; this may introduce a small systematic that has not been quantified or accounted for in the final uncertainty. The number of counts in the SR1 WS ROI was estimated as $1.4\pm0.4$.

\subsection{$^{37}$Ar}
\label{sec:Ar37}
 
$^{37}$Ar decays via electron capture to $^{37}$Cl which atomically relaxes via K-shell (2.82\,keV, 90.2$\%$), L-shell (0.270\,keV, 8.9$\%$), and M-shell (0.018\,keV, 0.9$\%$) cascades. $^{37}$Ar is produced via cosmogenic spallation during the storage and transit of the xenon from SLAC to SURF. The rate of $^{37}$Ar was estimated by calculating the exposure of the xenon to cosmic rays before it was transported underground, then correcting for the decay time before the search began \cite{LZ-Ar37}. The final prediction for $^{37}$Ar was 11\,nBq/kg, on 23 December 2021 (the start of SR1), producing an expected 96 events in the WS. This was allowed to float between 0 and 288 events in the statistical analysis as the uncertainty on the spallation yield is about a factor of three. 

Looking at the data in both the log$_{10}($S2$c$)-S1$c$ observable space and in reconstructed energy, there is a clear peak of events in the expected region for the $^{37}$Ar K-shell (see Figures~\ref{fig:WSEnergy} and \ref{fig:WSevents}). In the fits to the WS ROI, the best-fit number of $^{37}$Ar events was 52.5$^{+9.6}_{-8.9}$ (Section~\ref{sec:PLR}). A post-fit analysis was undertaken to understand whether the events in this region were consistent with the hypothesis of $^{37}$Ar decay. 

To perform this analysis, the 85 WS events that were within the 2$\sigma$ contour of the $^{37}$Ar location in the log$_{10}($S2$c$)-S1$c$ observable space were selected (see orange contour in Figure~\ref{fig:WSevents}). These events will include both $^{37}$Ar events with a decaying rate and other ER backgrounds, primarily from $^{214}$Pb, that were constant in time. The WS time period was divided into 13 bins: four before the calibration period in January 2022, seven between the calibration and the circulation event in March 2022, and two after the circulation event. Two models were fit to the data: one in which the rate of events in this region is constant in time, and one in which there is one component with a decay half-life of 35 days, consistent with that of $^{37}$Ar, and a flat component to account for the constant beta background. The model that is constant in time has a best fit of $1.32\pm0.14$ events per day. The model with exponential decay has a constant in time rate of $0.36\pm0.22$ events per day and a starting rate for the exponential model of $2.47\pm0.61$ events per day. 

\begin{figure}
    \centering
    \includegraphics[width=\columnwidth]{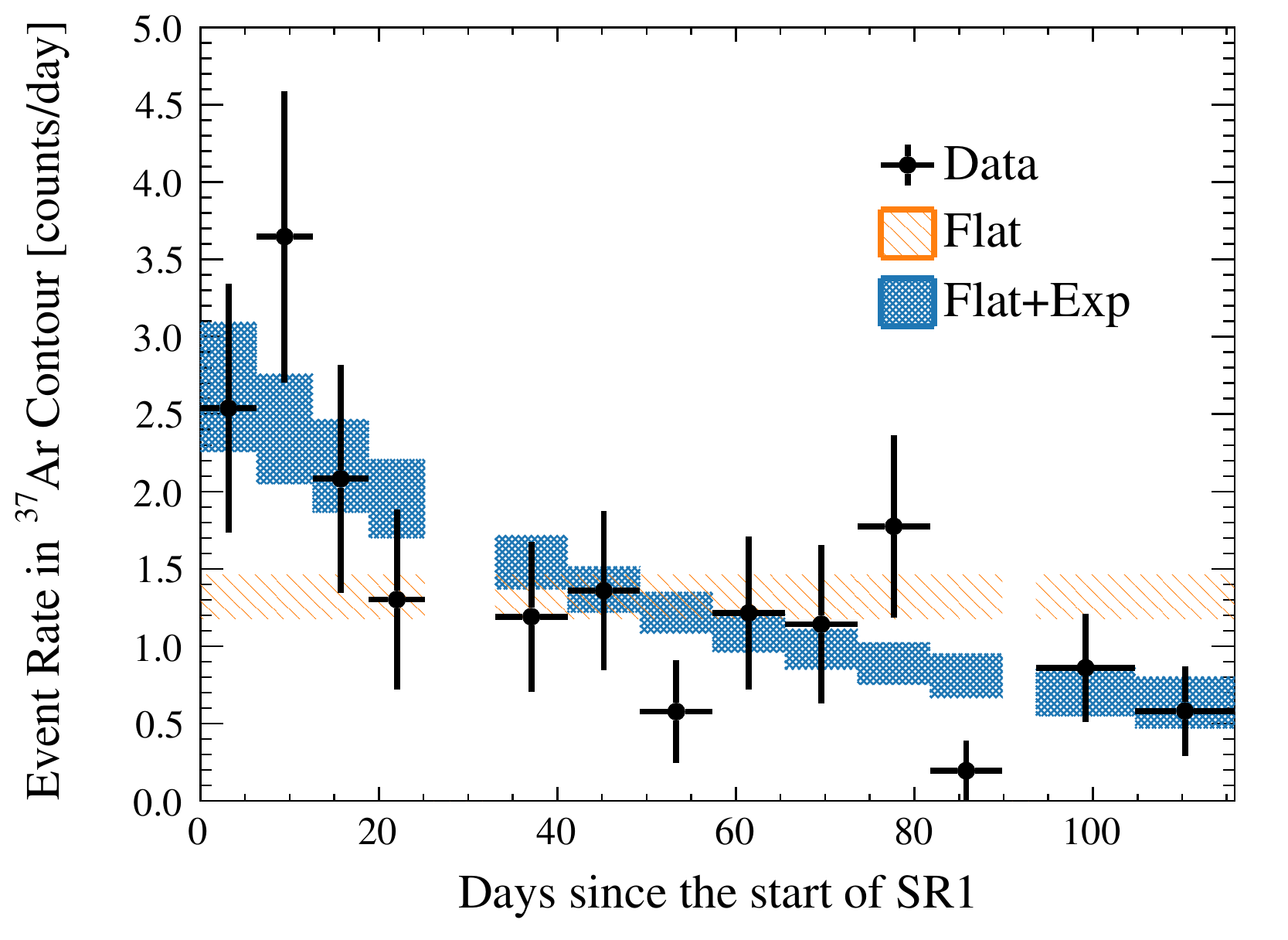}
    \caption{Time dependence of events in the 2$\sigma$ contour for $^{37}$Ar in log$_{10}($S2$c$)-S1$c$ space (see Figure~\ref{fig:WSevents}). Black points show the data. The orange band shows the best fit to a rate constant in time, including systematic uncertainty from the fit, and the blue band shows the best fit to a rate constant in time plus an exponential decay with a 35 day half-life. The data and predictions are corrected for the live time in each bin.}
    \label{fig:37Ardecay}
\end{figure}

Figure~\ref{fig:37Ardecay} shows the best fit with uncertainty for each of these two models. The constant in time model has a $\chi^2/\mbox{NDF} = 32.58/12$, which corresponds to a $p$-value of 0.0033, which is not consistent with the data. The model with exponential decay has a $\chi^2/\mbox{NDF} = 13.25/11$, which corresponds to a $p$-value of 0.43, therefore consistent with the data.

This analysis indicates that in SR1, $^{37}$Ar formed a significant part of the overall background, but is decaying away and will be substantially less prominent in future searches.

\subsection{Physics Backgrounds}
\label{sec:PhysicsBackgrounds}

Though interesting physics in their own right, $^{124}$Xe DEC, $^{136}$Xe $2\nu\beta\beta$ decay, solar and atmospheric neutrinos pose backgrounds in the WS ROI. Given the long half-life of both isotopes, these backgrounds were modeled as uniform in the TPC where each decay was assumed to be a single scatter ER without any interaction in the vetoes. The following subsections describe how each of the component rates were calculated, using literature values, and which energy spectra were used.

\subsubsection{$^{124}$Xe $2\nu$DEC}
\label{sec:Xe124ROI}
$^{124}$Xe $2\nu$DEC can occur with a combination of captures from K-, L-, M-, and N- shells:  the LL and LM decay modes are background contributors in the WIMP ROI, with total energies of 10.00\,keV and 5.98\,keV, respectively \cite{LARKINS1977311}. The first observation of $^{124}$Xe $2\nu$DEC was reported by the XENON1T experiment in Ref.~\cite{XENON1T-Xe124-2019}; the XENON Collaboration has since made further measurements to place its half-life at $(1.1\pm 0.2_{stat} \pm 0.1_{sys})\times 10^{22}$ years \cite{XENONnT-Xe124_Xe136-2022}, and also estimated the branching ratios of the LL mode as 1.4$\%$ and the LM mode as 0.8$\%$. With a natural abundance of $9.52\times10^{-4}$ \cite{XeAbundances}, the activity of $^{124}$Xe in the TPC was therefore predicted to be $8.72\pm2.44$\,nBq/kg. The log$_{10}($S2$c$)-S1$c$ response was evaluated using a modified NEST response that includes the measured perturbation of the L-shell contribution \cite{XELDA-Lshell}. The resulting number of counts in the SR1 WS ROI was predicted to be $5.0\pm1.4$.

\subsubsection{$^{136}$Xe $2\nu\beta\beta$ Decay}
\label{sec:Xe136ROI}

$^{136}$Xe $2\nu\beta\beta$ decay has been reported in multiple experiments \cite{PhysRevC.85.045504,Si:2022lyh}, and the half-life reported by EXO-200 is $2.165 \pm 0.016_{stat} \pm 0.059_{sys} \times 10^{21}$ years \cite{PhysRevC.89.015502}. Given an isotopic abundance of 8.9$\%$ \cite{XeAbundances}, the rate of $^{136}$Xe is 4.14\,$\pm$\,\SI{0.12}{\micro\becquerel/\kilogram} in the TPC. The underlying energy spectrum was also updated following the prescription in Ref.~\cite{Haselschwardt:2020iey}. The final counts in the SR1 WS ROI were evaluated as $15.1\pm2.4$.

\subsubsection{Solar Neutrinos}
\label{sec:NeutrinoROIBkgs}

Low-energy ER events are induced by solar neutrinos undergoing electroweak interactions with the active liquid xenon volume. The spectra are dominated by $pp$ chain neutrinos, with small contributions from $^{7}$Be and CNO neutrinos. The low-energy portion of the recoil spectra was calculated using the relativistic random phase approximation (RRPA) to account for atomic binding effects, as described in Ref.~\cite{RRPA_Chen:2016eab}. The recoil spectra in the WIMP ROI are approximately flat and the rate was calculated based on measurements from Refs.~\cite{Borexino_PhysRevD.100.082004,SNO_PhysRevC.88.025501,SSM_Vinyoles:2016djt}. $27.1\pm1.6$ solar neutrino ER events in the SR1 WS ROI were expected.

Solar neutrinos can additionally interact via coherent elastic neutrino-nucleus scattering (CE$\nu$NS) in the case of $^{8}$B and, to a much lesser extent, \textit{hep} solar, diffuse supernova (DSN) and atmospheric neutrinos \cite{COHERENT:2017ipa}. The prediction for the CE$\nu$NS rate is described in Refs.~\cite{Borexino_PhysRevD.100.082004,SNO_PhysRevC.88.025501,SSM_Vinyoles:2016djt}, but was heavily suppressed in the SR1 WIMP ROI due to the S2$_\text{raw}>600$\,phd threshold. Only the dominant $^{8}$B contribution was considered, giving a prediction of $0.14\pm0.01$ events in the SR1 WS ROI, with only a small flux uncertainty.

\subsection{$^{127}$Xe}
\label{sec:Xe127}

Cosmogenically-activated $^{127}$Xe was present during SR1 at an exposure-averaged rate of \SI{32.8}{\micro\becquerel/\kilogram}, as derived by the extrapolation of the activity determined at the start of SR1 through energy-spectrum fitting (Table~\ref{tab:InnerTonneFits}, Section~\ref{sec:BetaFit}). $^{127}$Xe can be problematic for the WS if the de-excitation gamma ray(s) following the electron capture are not detected. This can happen when the $^{127}$Xe atoms are located near the edges of the active liquid xenon volume, where gamma rays can escape the TPC, leaving behind semi-naked vacancy cascades that can fall into the WIMP ROI. This would be the case for 5.2\,keV L- and 1.1\,keV M-shell cascades, which account for 13.1\,$\%$ and 2.9\,$\%$ of all $^{127}$Xe electron captures, respectively \cite{LUX-Xe127}. 

Most of these semi-naked events can be vetoed by the Skin and OD systems through detection of the gamma rays that elude the TPC. Semi-naked K-shell events were used to determine the veto efficiency in SR1, being outside of the WIMP ROI and occurring at a sufficiently high rate to form a useful sideband. As only those L- and M-shell events which survive the FV selection were of interest to the WS, the FV cut was applied to the K-shell events to ensure no extra systematic was introduced in calculating the veto efficiency. Candidate K-shell events were identified via a selection in the observable log$_{10}$(S2$c$)-S1$c$ space. Electron capture decays of $^{125}$I and $^{124}$Xe, specifically the (35.5+4)\,keV $^{125}$I gamma ray plus L-shell cascade and K- plus L-, M- or N- shell contributions of $^{124}$Xe $2\nu$DEC, can leak into this selection. Prior to fitting the veto-untagged populations in reconstructed energy space, the $^{124}$Xe rate was fixed based on the literature values discussed in Section~\ref{sec:Xe124ROI}. The fit was also performed for veto-tagged events; the ratio of the semi-naked $^{127}$Xe K-shell events determined in the two fits was used to establish the veto efficiency in the fiducial volume at $78\pm5$\%.

The fraction of semi-naked L- and M-shell events in the fiducial volume and WIMP ROI was determined using the Monte Carlo simulations described in Section ~\ref{sec:Sims} and cross-checked against data. The rate of $^{127}$Xe relevant to the WIMP analysis was ultimately calculated from the product of this factor with the SR1-averaged activity and the veto inefficiency. The SR1 WS counts were thus estimated to $9.2\pm0.8$ events. For the $^{127}$Xe PDF used in the profile likelihood ratio analysis, the L-shell cascade response was simulated with modified recombination fractions, as empirically measured in Ref.~\cite{XELDA-Lshell}.

\subsection{Accidental Backgrounds}
\label{sec:Accidentals}

An instrumental background was induced by the accidental coincidences of pulses identified as S1s and S2s that have unrelated origins, which were classified as SS events by the LZ event reconstruction framework. These so-called accidental events can mimic standard S1-S2 pairs at a variety of energies, but predominantly those of low-energy recoils in the WIMP ROI.

Pulses that contribute to accidental events can originate from several regions in the detector. S1s unpaired with S2s can arise from charge-insensitive regions of the TPC, such as the reverse field region below the cathode. S2s unpaired with S1s can come from regions with poor light collection or from events where the S1 is below detection threshold. Additionally, lone S2s can occur for interactions towards the top of the detector, where the short drift time means the S1 is subsumed within or cannot be separated from the S2 pulse. Non-xenon processes can also yield pulses that can be misidentified as S1s and S2s. For example, Cherenkov light in the quartz window of a PMT or dark noise pile-up from PMTs can appear as S1s, whilst electron emission from the cathode or gate electrodes could be tagged as S2s.

\begin{figure}[!ht]
    \centering
    \includegraphics[width=\columnwidth]{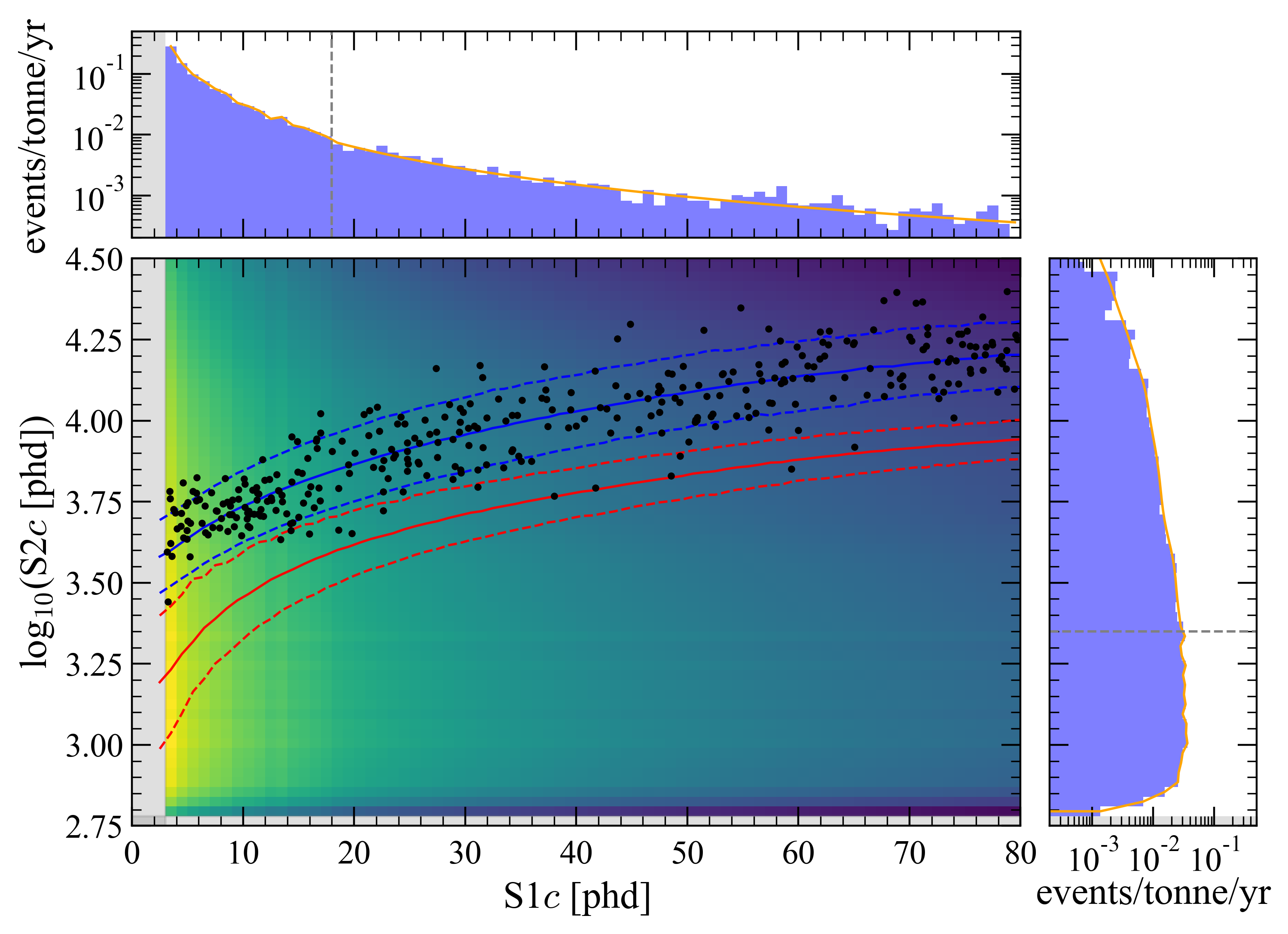}
    \caption{The accidentals PDF normalized to events/tonne/yr. The 10\% and 90\% quantiles (dashed) as well as the median (solid) of the ER and NR bands, as reported in Ref.~\cite{LZ-SR1result}, are shown in blue and red, respectively. SR1 WS events remaining after all data selections are also shown (black dots). The regions outside the WS ROI are marked with a shaded gray area. The number of predicted accidental events in the entire ROI is 1.2 (0.2 inside the NR band). The top and right panels show the projections on each axis of the ACS events surviving all analysis cuts, where the distribution falls off at small $\text{log}_{10}(\text{S2}c)$ due to the applied 600 phd lower bound in raw S2 pulse size. The orange lines represent the functions that were used to build the two-dimensional PDF. Regions with larger data fluctuations, starting from the dashed gray lines, were smoothed out.}
    \label{fig:accidentals_pdf}
\end{figure}

Analysis cuts were developed to specifically target many of the aforementioned sources of pulses classified as S1s and S2s. The cuts used properties of the individual pulses, such as their shape or the observed hit patterns of the light on the PMTs, to distinguish them from typical S1s and S2s. A subset of cuts also exploited the physical relationship of certain parameters with the drift time established for the given S1-S2 pair. In particular, the S1 TBA is expected to decrease with increasing drift time, as more S1 light from events lower in the TPC is seen in the bottom PMT array. Also, the width of an S2 pulse is expected to increase with drift time, since the ionization electron cloud has more time to diffuse as it drifts towards the extraction region. Further cuts were developed to target events occurring near the liquid surface, which showed a characteristic narrow width and square shape, as well as events happening between the liquid surface and the anode, in which case the S1 and S2 pulse were often merged resulting in an atypical, elongated S2 pulse. 

A useful auxiliary dataset to study the accidental background was formed from Unphysical Drift Time (UDT) events. UDT events have reported drift time exceeding the maximum value of \SI{951}{\us} measured in SR1, and must therefore be formed by S1-S2 pairs that were not due to standard single interactions in the xenon and/or were not physically correlated, which would indicate that these events must be of accidental origin. Several statistical tests were conducted to test the independence of the S1 and S2 variables in the UDT population and no significant correlation was found, confirming that UDT events were suitable for investigating accidental coincidences. However, the number of UDT events in SR1 was too small to comprehensively map out the distribution of accidental events in the WIMP ROI, especially after applying accidental-specific analysis cuts.

To model the accidental background, a data-driven approach was taken, combining pulses at the waveform level to manufacture artificial accidental events, called \dquotes{Accidental ChopStitch} (ACS) events. S1 and S2 pulses were selected from Other classified events, as opposed to those classified as SS or MS. These events were required to pass the SR1 live time vetoes and a subset of the WS data quality cuts designed to remove noise-prone events. The UDT population that persists after WS cuts was dominated by events where a trigger was observed on the S2. Therefore, candidate S1s to create ACS events were required to be in the pre-trigger region, and candidate S2s required to be straddling the trigger point of their respective events. ACS events were manufactured splicing together the period before the start of the S2 from the candidate S1 event with the period including and after the S2 from the candidate S2 event. The pulse environment in the vicinity of each pulse was therefore preserved, which can affect the classification of the combined event. No further criteria were applied in the selection of the S1 and S2 pulses, and the independence of the S1 and S2 pulse areas in the resultant ACS events was verified via several statistical tests.

The way ACS events are manufactured ensures the eventual decision of whether the resultant S1-S2 pair forms a SS event lies entirely with the LZ event classification algorithm. One of the main advantages ACS events offer is that they can be processed as real data through the LZ event reconstruction framework, reducing systematic uncertainties associated with finding and classifying accidental events.

To investigate the agreement between the UDT population and the manufactured ACS population, the spectra of the UDT and ACS populations in several dimensions of interest were compared (e.g.~pulse size, S1 TBA, drift time). A KS test in each case indicated that the two distributions were not distinct enough to fail the test, reinforcing the idea that ACS events were a good proxy for accidentals. 

30 million ACS events were generated, with only approximately 22,000 events passing all the analysis cuts. This number of events were not sufficient to produce a smooth PDF across the WS ROI. Instead, the S1-S2 distribution was projected into its corresponding axes and an interpolating function was found for those regions at large pulse areas where the number of events was low. The two-dimensional PDF was constructed by taking the outer product of the smoothed out versions of the S1 and S2 projections. The resulting PDF is shown in Figure~\ref{fig:accidentals_pdf}.

The PDF was normalized using an independent calculation of the accidentals rate. The expected number of accidental events in the WS ROI was calculated by multiplying two quantities: the observed number of UDT events after applying a basic selection of data quality cuts that exclude periods of elevated TPC activity and electronics interference but before applying any of the pulse-based cuts (310 events), and the rejection efficiency evaluated on the ACS population after applying all the analysis cuts (99.6\% efficiency). This method combines the best characteristics of each data sample: the UDT population, which is anchored on an experimental measurement, and the ACS population, which contains an arbitrarily large number of manufactured events. The predicted number of accidental events in the SR1 livetime following this method was $1.2\pm0.3$. Both statistical and systematic uncertainties were accounted for in this calculation, with the dominant ones being the systematic uncertainties arising from the difference observed when S1 and S2 shape cuts were applied to either UDT or ACS data. Nevertheless, an overall good agreement was observed when the full list of analysis cuts was applied to both the UDT and ACS populations, resulting in a similar rejection efficiency.

\begin{figure}[h]
    \centering
    \includegraphics[width=\columnwidth]{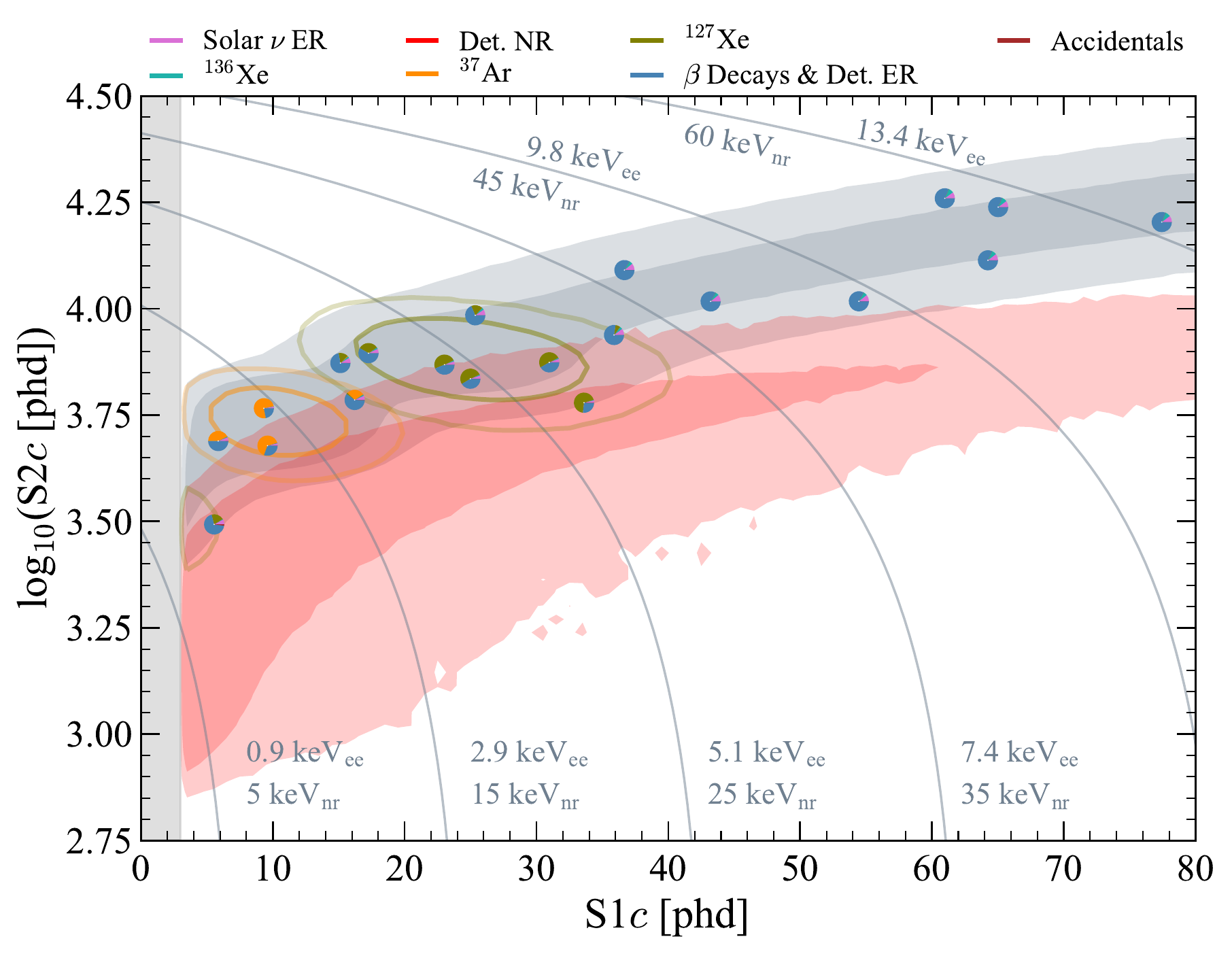}
    \caption{Events that passed all WS cuts, but were tagged by the OD (i.e. fail the WS OD veto cut) are shown in log$_{10}($S2$c$)-S1$c$ space. Each data point is represented as a pie chart, with sectors representing the likelihood it originated from the given background. 1$\sigma$ and 2$\sigma$ contours for background and signal (neutrons in this case) are overlaid.}
    \label{fig:ODTaggedEvents}
\end{figure}

\begin{table}[htbp]
    \caption{Expected and fit values for each contribution in the fit to the OD-tagged event sample. The detector NR population was left unconstrained.}
    \centering
    \setlength{\tabcolsep}{7pt}
    \begin{tabular}{ccc}
    \tabularnewline
    \hline
    \hline
    Source & Expected Events & Fit Result\tabularnewline
    \hline
    Solar $\nu$ ER & 1.44 $\pm$ 0.03  & 1.43 $\pm$ 0.03  \tabularnewline
    Detector neutrons & 0.8 & 0.0$^{+0.8}$ \tabularnewline
    \ce{^{37}Ar} & 2.9 $\pm$ 0.5 & 2.8 $\pm$ 0.5 \tabularnewline
    \ce{^{136}Xe} &	0.79 $\pm$ 0.12 & 0.79 $\pm$ 0.12 \tabularnewline
    \ce{^{127}Xe} & 1.6 $\pm$ 0.2 & 1.6 $\pm$ 0.2 \tabularnewline
    $\beta$ decays + Det. ER & 10.7 $\pm$ 2.6 & 11.3 $\pm$ 2.2 \tabularnewline
    Accidentals	& 0.09 $\pm$ 0.03 & 0.10 $\pm$ 0.03 \tabularnewline
    \hline
    Total & 18.2 $\pm$ 2.7 & 18.1 $\pm$ 2.4 \tabularnewline
    \hline
    \hline
    \end{tabular}\\
        \label{tab:neutronFit}
\end{table}

\subsection{Neutrons}
\label{sec:NeutronConstraint}
The constraint on the neutron background level for the WS was derived using an auxiliary fit to events which failed the OD veto cut, but which otherwise passed all selection cuts. The fit was performed analogously to that for the WS (Section~\ref{sec:PLR}), with the expected rate of each background set to 5\% of that determined for the WS sample. This 5\% accounted for the chance of an accidental OD-TPC coincidence that would cause the event to fail the OD veto cut, in which an unrelated, above-threshold OD pulse would have occurred within the \SI{1200}{\us} veto window following the TPC signal. An exception was made for $^{127}$Xe, where OD coincidences were anticipated, and thus the expected value was set to $1.6\pm0.2$ based on its veto efficiency derived in Section~\ref{sec:Xe127}. The OD-tagged events are shown in Figure~\ref{fig:ODTaggedEvents}. The best-fit number of neutrons in this sample was $0.0^{+0.8}$, as illustrated in the fit results in Table~\ref{tab:neutronFit}. The resulting shape of the log-likelihood profiled in the amount of neutron background was well-characterized by a fourth-order polynomial and was used for the shape of the constraint on the neutron population in the WS fit. The predicted neutron background in the WIMP fit region, derived from the same profiled log-likelihood, was $0.0^{+0.2}$.

\begin{figure}
    \centering
    \includegraphics[width=\columnwidth]{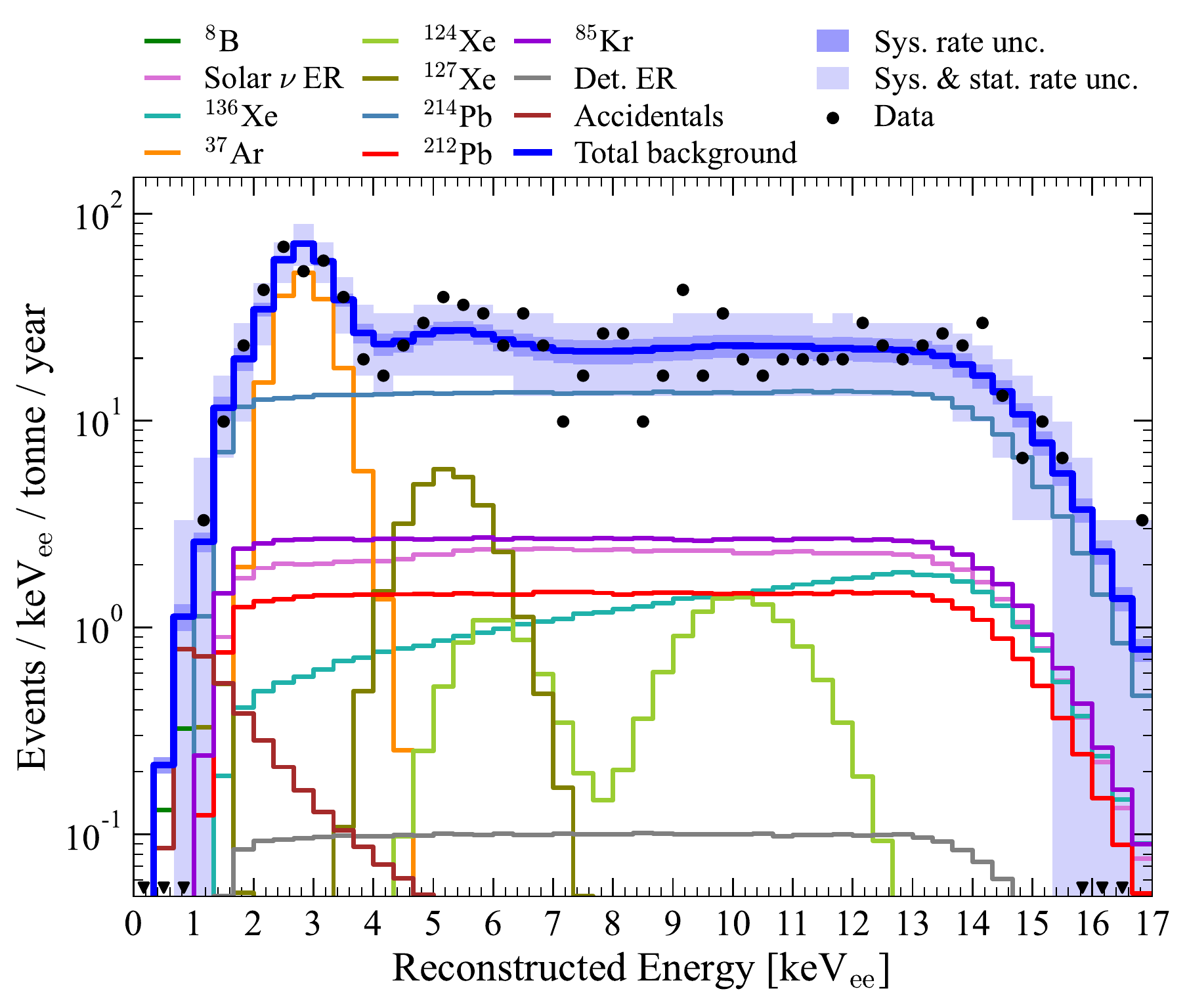}
    \caption{Background model before fitting to the SR1 data (except for the $^{37}$Ar component, for which the post-fit normalization is used). The total model is shown in dark blue, and SR1 data after all WS cuts have been applied are denoted by the black points. This represents a background event rate of $(6.3 \pm 0.5) \times 10^{-5}$ events/keV$_{ee}$/kg/day.}
    \label{fig:WSEnergy}
\end{figure}

A secondary calculation was made using the results in Section~\ref{sec:NeutronCounting}. To convert from the ten observed MS events to the number of SS events that pass all cuts, several factors needed to be applied. Systematic effects were difficult to quantify for the initial neutron event observation and for some of these factors, thus uncertainties are not quoted here. The 49\% OD veto efficiency established for an OD threshold of 400\,phe and a coincidence window of \SI{400}{\us} was unfolded and a 20\% OD veto inefficiency for the WS veto cut of 37.6\,phe threshold and \SI{1200}{\us} TPC S1-OD time separation was factored in to give the number of OD-untagged MS counts. Then a simulation-based MS:SS ratio of 2.3:1 was used to convert to SS counts. None of the calibration energy spectra closely approximate those of radiogenic neutrons, thus the factor was derived from simulations as it was expected to be dependent on energy. Nevertheless, the number compares favourably to those determined from DD (2.0:1) and AmLi (1.3:1) calibrations, which were not as high in energy as the radiogenic neutrons on average. Finally, the survival fraction for FV and ROI cuts was calculated from simulations and applied. After all these considerations, 0.29 events were estimated for the WS events surviving the OD veto cut, and 1.1 events were estimated for the number of OD-tagged WS events. These values are comparable to the $0.0^{+0.2}$ WS and $0.0^{+0.8}$ OD-tagged fit-derived neutron background contributions, respectively.

\begin{table}
    \caption{Number of events from background components in the WIMP ROI in the $330\pm12$ tonne-days SR1 exposure. The middle column shows the predicted number of events with uncertainties as described in Sections~\ref{sec:BetaAndGamma} to \ref{sec:NeutronConstraint}. The uncertainties were used as constraint terms in a combined fit of the background model plus a \SI{30}{GeV/c\squared} WIMP signal to the selected data, the results of which are shown in the right column. $^{37}$Ar and detector neutrons used non-Gaussian prior constraints and are totaled separately. Values at zero have no lower uncertainty.}
    \label{tab:bkgds}
    \centering
    \begin{tabular}{>{\centering}p{3.3cm}>{\centering}p{2.38cm}>{\centering}p{2.3cm}}
    \tabularnewline
    \hline
    \hline
    Source & Expected Events & Fit Result\tabularnewline
    \hline
    \ce{^{214}Pb} & \centering 164 $\pm$ 35\phantom{0} &\centering - \tabularnewline
    \ce{^{212}Pb} & \centering 18 $\pm$ 5\phantom{0} &\centering - \tabularnewline
    \ce{^{85}Kr} & \centering 32 $\pm$ 5\phantom{0} &\centering - \tabularnewline
    Det. ER & \centering 1.4 $\pm$ 0.4 &\centering -  \tabularnewline
    \hline
    $\beta$ decays + Det. ER &\centering 215 $\pm$ 36\phantom{0} &\centering 222 $\pm$ 16\phantom{0}  \tabularnewline
    \hline
    $\nu$ ER& 27.1 $\pm$ 1.6\phantom{0} & 27.2 $\pm$ 1.6\phantom{0} \tabularnewline
    \XeOneTwoSeven & 9.2 $\pm$ 0.8 & 9.3 $\pm$ 0.8 \tabularnewline 
    \ce{^{124}Xe}& 5.0 $\pm$ 1.4 & 5.2 $\pm$ 1.4 \tabularnewline
    \ce{^{136}Xe} &	15.1 $\pm$ 2.4\phantom{0} & 15.2 $\pm$ 2.4\phantom{0} \tabularnewline
    $^8$B CE$\nu$NS & 0.14 $\pm$ 0.01 & 0.15 $\pm$ 0.01 \tabularnewline
    Accidentals	& 1.2 $\pm$ 0.3 & 1.2 $\pm$ 0.3 \tabularnewline
    \hline
    \rule{0pt}{1.0ex} Subtotal & 273 $\pm$ 36\phantom{0} 
    & 280 $\pm$ 16\phantom{0} \tabularnewline
    \hline \tabularnewline[-2.2ex]
    \ce{^{37}Ar} & [0, 288] & $52.5^{+9.6}_{-8.9}$\phantom{0} \tabularnewline[0.25ex]
    Detector neutrons &	$0.0^{+0.2 }$ & $0.0^{+0.2}$ \tabularnewline[0.25ex]
    \SI{30}{GeV/c\squared} WIMP &\centering -- &\centering $0.0^{+0.6}$ \tabularnewline[0.25ex]
    \hline
    Total & -- & 333 $\pm$ 17\phantom{0} \tabularnewline
    \hline
    \hline
    \end{tabular}
\end{table}

\begin{figure*}[!ht]
  \centering
  \subfloat{\includegraphics[width=.49\textwidth]{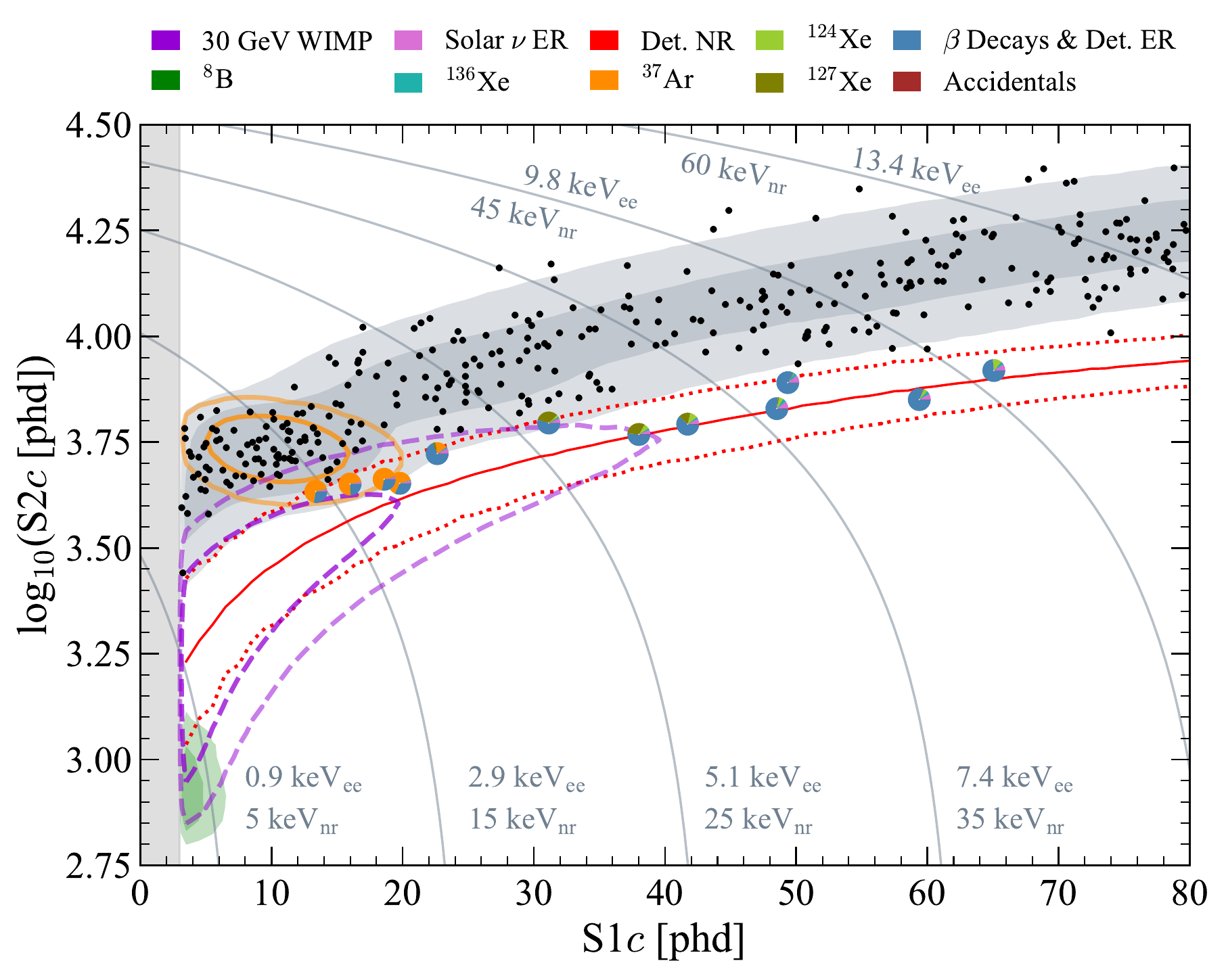}  \label{fig:WSevents_s2vsS1}}
  \hfill
  \subfloat{\includegraphics[width=.49\textwidth]{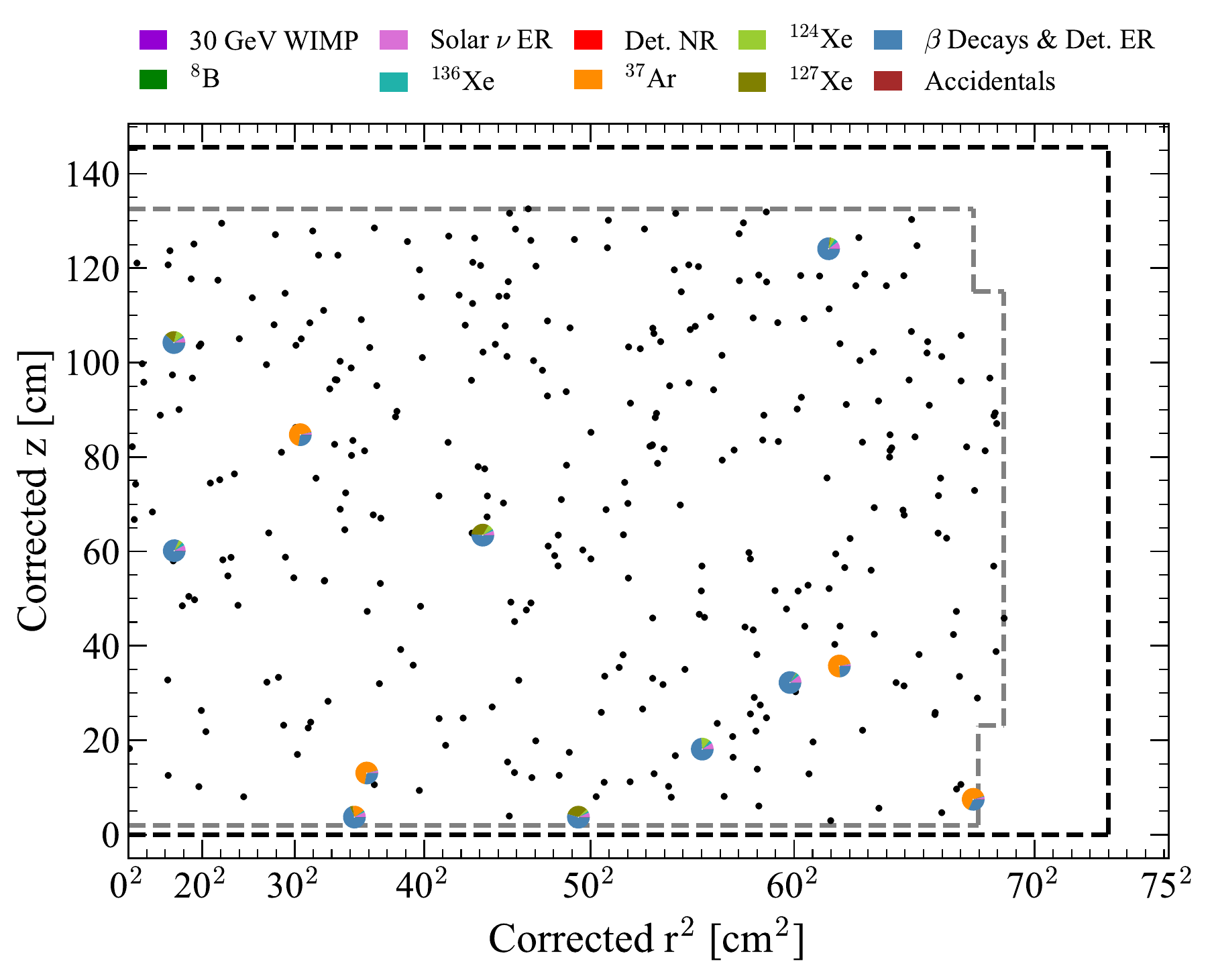} \label{fig:WSevents_zvsR2}}
  \caption{\emph{Left:} Low-energy data after all data quality and physics cuts in log$_{10}($S2$c$)-S1$c$ space. Contours enclose 1$\sigma$ and 2$\sigma$ of the best-fit background model (shaded gray), the $^{37}$Ar component (orange ellipses), a 30\,GeV/c$^2$ WIMP (purple dashed lines), and $^{8}$B solar neutrinos (shaded green regions). The solid red line shows the NR median, and the red dotted lines indicate the 10\% and 90\% quantiles. \emph{Right:} Corrected \textit{z} and \textit{r$^{2}$} positions of the same data. Dashed black lines outline the active liquid xenon volume and dashed gray lines represent the fiducial volume. In both figures, events falling below the 2$\sigma$ contour of the best-fit background model are shown as pie charts for which the size of each wedge is determined by the relative weight of each of the background components in the fit.}
  \label{fig:WSevents}
\end{figure*}

\subsection{Background Expectations Summary and Fit}
\label{sec:PLR}
The complete background expectations, as per the discussions in Sections~\ref{sec:BetaAndGamma} to \ref{sec:NeutronConstraint}, can be found in Table~\ref{tab:bkgds}. Figure~\ref{fig:WSEnergy} illustrates these in reconstructed energy space within the WIMP ROI, using the simulated events for each component, scaling each as per its assessed contribution in the table. This is with the exception of $^{37}$Ar, which did not have a tight prior constraint (Section~\ref{sec:Ar37}), and for which the normalization derived from the following fit procedure was used.

Events in the WS ROI which pass all data quality and physics cuts were fitted using an extended, unbinned likelihood containing both signal (WIMP) and background components,
\begin{equation}
\label{eq:likelihood}
\begin{split}
\mathcal{L}(\mu_{s},\pmb{\theta}) = & \text{Pois}(N_{0}|\mu_{\rm tot}) \\ & \times \prod_{e=1}^{N_{0}} \frac{1}{\mu_{\rm tot}} \left( \mu_{s}f_{s}(\pmb{x_{e}}) + \sum_{b=1}^{N_{b}} \mu_{b}f_{b}(\pmb{x_{e}}) \right) \\ & \times \prod_{b=1}^{N_{b}} f_{b}(\mu_{b}|\nu_{b}) \, ,
\end{split}
\end{equation}
where $\mu_{\rm tot}=\mu_{s}+\sum_{b=1}^{N_{b}} \mu_{b}$ is the sum of signal and background levels and $e$ is an index which runs up to the total number of observed events, $N_{0}$. Both signal ($f_{s}$) and background PDFs ($f_{b}$) are functions of the analysis parameters $\pmb{x_{e}} = \left(\text{S1}c, \text{log}_{10}(\text{S2}c) \right)$. The set of nuisance parameters $\pmb{\theta}$ is the set of counts for each background component $\{ \mu_{b} \}$. Constraint functions, $f_{b}(\mu_{b}|\nu_{b})$, limit the value of each nuisance parameter to that expected from the auxiliary measurements presented in this paper ($\nu_{b}$).

For all components, except the $^{37}$Ar and neutron backgrounds, the constraints were Gaussian with standard deviation corresponding to the systematic uncertainty on the expectation of each background. The $^{37}$Ar background was constrained using a uniform distribution between 0 and 288 events, the latter being three times the expected number of events in the exposure based on predictions for its production in the LZ xenon payload while on the surface~\cite{LZ-Ar37} (Section~\ref{sec:Ar37}). A constraint on the number of neutron NR events was derived from a fit to events tagged in the OD, as described in Section~\ref{sec:NeutronConstraint}. The results of the fit are listed in Table~\ref{tab:bkgds}, and all 335 events passing data quality and physics cuts are shown in Fig.~\ref{fig:WSevents}. Pie charts are used for the events in the NR band, showing how they have been attributed to different background components as a result of the fitting.

It can be seen from Table~\ref{tab:bkgds} that, excluding the special case of $^{37}$Ar, the likelihood fit does not provide better constraints for the background components than our pre-fit assessments in the vast majority of cases. In other words, our background model, determined without using the WS data, is consistent with the fit to the WS data, with better precision than the WS data alone can provide.

\section{Conclusions}
\label{sec:Conclusions}
A backgrounds model for LZ was developed with analysis of data from the first science run, and was successfully employed in the inaugural WIMP results reported by the experiment in Ref.~\cite{LZ-SR1result}. The pre-fit model agrees well with the observed data in the WIMP ROI (Figure~\ref{fig:WSEnergy}). Sources outside the WIMP ROI were well-characterized, which enables their inferred activities to inform additional physics searches in a broader ROI. Fitting of the gamma-ray sources and radon alpha decays was achieved to good precision; gamma-ray activities were found to be compatible with assay expectations, whereas radon levels were found to be higher than expected. The results inform strategies for radon emanation and assay measurements for future experiments. 

The model is comprehensive across a wide range of energies, and can be easily adapted for other physics searches. The backgrounds in the higher energy regions up and beyond the ROI relevant for $^{136}$Xe $0\nu\beta\beta$ have been characterized, ready for blinded searches with future science runs. For WIMPs and low-energy searches, these background results can be used as the basis for understanding how to optimize the detector conditions for later science data-taking; for instance, investigating how to push analysis thresholds whilst maintaining a workable accidental coincidence rate. All three detectors were leveraged in the determination of contributions to the model, with the assessed veto performance in the OD and Skin being instrumental to estimates for neutrons and $^{127}$Xe. Work is ongoing to understand the background events observed in the veto detectors themselves, which can in turn further inform the TPC model.

The background rate in the WIMP ROI was established as $(6.3\pm0.5)\times10^{-5}$ events/keV$_{ee}$/kg/day: this represents a 57 times reduction over the background rate of $(3.6\pm0.4)\times10^{-3}$ events/keV$_{ee}$/kg/day reported by LUX after their WS criteria were applied in Ref.~\cite{LUX-BG}. This rate is likely to improve further as the components in the model evolve with time. The cosmogenically-activated xenon and $^{37}$Ar will decay to subdominant levels. On the other hand, the state of the detector becomes more variable with longer exposures, which could, for example, lead to enhancement of sources contributing to accidental coincidences. The WS ROI definition, cuts, and FV may also change in the future, which would alter the background profile in consideration for a next WIMP analysis. A more sophisticated profile-likelihood ratio analysis involving more parameters, such as time dependence, could be developed to better utilize the background information detailed in this paper. Analyses presented here, such as the radon alpha movement studies, demonstrate that position dependence is also viable for a next physics analysis. 

\begin{acknowledgments}
The research supporting this work took place in part at SURF in Lead, South Dakota. Funding for this work is supported by the U.S. Department of Energy, Office of Science, Office of High Energy Physics under Contract Numbers DE-AC02-05CH11231, DE-SC0020216, DE-SC0012704, DE-SC0010010, DE-AC02-07CH11359, DE-SC0012161, DE-SC0015910, DE-SC0014223, DE-SC0010813, DE-SC0009999, DE-NA0003180, DE-SC0011702, DE-SC0010072, DE-SC0015708, DE-SC0006605, DE-SC0008475, DE-SC0019193, DE-FG02-10ER46709, UW PRJ82AJ, DE-SC0013542, DE-AC02-76SF00515, DE-SC0018982, DE-SC0019066, DE-SC0015535, DE-SC0019319, DE-AC52-07NA27344, \& DOE-SC0012447.	This research was also supported by U.S. National Science Foundation (NSF); the UKRI’s Science \& Technology Facilities Council under award numbers ST/M003744/1, ST/M003655/1, ST/M003639/1, ST/M003604/1, ST/M003779/1, ST/M003469/1, ST/M003981/1, ST/N000250/1, ST/N000269/1, ST/N000242/1, ST/N000331/1, ST/N000447/1, ST/N000277/1, ST/N000285/1, ST/S000801/1, ST/S000828/1, ST/S000739/1, ST/S000879/1, ST/S000933/1, ST/S000844/1, ST/S000747/1, ST/S000666/1, ST/R003181/1; Portuguese Foundation for Science and Technology (FCT) under award numbers PTDC/FIS-PAR/2831/2020; the Institute for Basic Science, Korea (budget number IBS-R016-D1). We acknowledge additional support from the STFC Boulby Underground Laboratory in the U.K., the GridPP~\cite{faulkner2005gridpp,britton2009gridpp}  and IRIS Collaborations, in particular at Imperial College London and additional support by the University College London (UCL) Cosmoparticle Initiative. We acknowledge additional support from the Center for the Fundamental Physics of the Universe, Brown University. K.T. Lesko acknowledges the support of Brasenose College and Oxford University. The LZ Collaboration acknowledges key contributions of Dr. Sidney Cahn, Yale University, in the production of calibration sources. This research used resources of the National Energy Research Scientific Computing Center, a DOE Office of Science User Facility supported by the Office of Science of the U.S. Department of Energy under Contract No. DE-AC02-05CH11231. We gratefully acknowledge support from GitLab through its GitLab for Education Program. The University of Edinburgh is a charitable body, registered in Scotland, with the registration number SC005336. The assistance of SURF and its personnel in providing physical access and general logistical and technical support is acknowledged. We acknowledge the South Dakota Governor's office, the South Dakota Community Foundation, the South Dakota State University Foundation, and the University of South Dakota Foundation for use of xenon. We also acknowledge the University of Alabama for providing xenon.  For the purpose of open access, the authors have applied a Creative Commons Attribution (CC BY) licence to any Author Accepted Manuscript version arising from this submission.

\end{acknowledgments}


%

\end{document}